%% file: Thesis.tex
\documentclass[11pt,final]{ucthesis}

\input{MyPackages.tex}
\input{MyMacros.tex}

\usepackage{url}

\renewcommand{\phi}{\varphi}

\begin{document}

\title{Nucleation, Growth, and Coarsening --- A Global View on Aggregation}
\author{Joseph Farjoun}
\degreeyear{2006}
\degreeterm{Spring}
\degree{Doctor of Philosophy}
\chair{Professor John C. Neu}
\othermembers{Professor Lawrence C. Evans\\
Associate Professor Kristofer S. J. Pister}
\numberofmembers{3}
\prevdegrees{B.S. (Hebrew University of Jerusalem, Israel) 2000}
\field{Mathematics}
\campus{Berkeley}

\maketitle
\approvalpage
\copyrightpage

\begin{abstract}
We present a new model of homogeneous aggregation that contains the
  essential physical ideas of the classical 
  predecessors, the Becker-D\"oring and Lifshitz-Slyovoz models.
  These classical models, which give different predictions, are
  asymptotic limits of the new model at small (BD) and large (LS)
  cluster sizes.  
  Since the new theory is valid for large and small clusters, it allows
  for a complete description of the nucleation  process; one that can
  predict the creation of super-critical clusters at the Zeldovich
  nucleation rate, and the diffusion limited  growth of large clusters
  during coarsening. 
  By retaining the physically valid ingredients from both models, we
  explain the seeming incompatibilities and arbitrary choices of the
  classical models. 

We solve the equations of our new model
asymptotically in the small super-saturation limit.
The solution exhibits three successive `eras': nucleation,
growth, and coarsening,
each with its specific scales of time and cluster size.
During the \emph{nucleation} era, the bulk of the
clusters are formed by favorable fluctuations over a free energy
barrier, according to the analysis by  Zeldovich.
The free energy barrier increases as more clusters are formed, and this
signals the beginning of the \emph{growth} era: no new clusters are
created, and the expansion of the existing ones continues.
The growth of the clusters slows down when the reservoir of monomers
that fuels it is sufficiently depleted. 
This signals the onset of the final \emph{coarsening} era.
This is a competitive attrition process, of smaller clusters
dissolving and fueling the further growth of the larger survivors.
By resolving the preceding creation and growth eras, our analysis
gives explicitly the characteristic time and cluster size of the
coarsening era, and a unique selection of the long time, self-similar
cluster size distribution.
\abstractsignature
\end{abstract}

\begin{frontmatter}

\tableofcontents
\listoffigures
\begin{acknowledgements}
I want to thank my advisor, John Neu, for long discussions, helpful
comments and guidance. 
\end{acknowledgements}

\end{frontmatter}
\chapter{A Global Model}

\section{Introduction}
Nucleation refers to the aggregation of identical particles (monomers) into clusters. 
Its universality throughout physics, chemistry and biology is well known. 
References \cite{KGT83}, \cite{KELT91}, \cite{NB00}, \cite{GWS01},
\cite{ISRAELACHVILI91}, \cite{NCB02}, \cite{LS61}, \cite{XH91},
\cite{MG96}, \cite{GNON03} provide a lineup of the `usual' (and some unusual) suspects.
Also well known are the long-standing challenges that aggregation poses to modeling.
Two classical models of aggregation due to Becker-D\"oring (BD) \cite{BD35}, and Lipshitz-Slyozov (LS) \cite{LS61} are incomplete and mutually inconsistent.

In BD, clusters exchange particles with the surrounding monomer bath by a `surface reaction', and it is assumed that the monomer bath around the clusters has uniform concentration.
This is only possible with infinite diffusivity of monomers. 
While this description is asymptotically accurate for sufficiently small clusters,
 the uptake of monomers by large clusters is strongly controlled by  the diffusivity.
LS describes cluster growth and shrinkage controlled by  diffusion of monomers.
In LS, the monomer concentration at the surface of a cluster is a prescribed function of the local curvature, generally different from the `background' concentration, far from clusters.
Hence, monomer concentration about a large cluster is nonuniform, and there is diffusive transport of monomer into or away from the cluster.
This physics of LS leads to a prediction for cluster growth that disagrees with BD\@.
Furthermore, LS is `incomplete', in that it does not describe the initial creation of clusters from pure monomer.
While it is generally accepted that BD is a model for small clusters, and
  LS  for large, several questions remain.
How to interpolate between the two models?
What is the characteristic size that separates `large' from `small'? 
What physics governs the growth in the intermediate scale?
What globally valid model encompasses the whole evolution of clusters,
  from an initial state of pure monomer to the asymptotic self-similar distribution of large cluster sizes?

The current chapter presents a new model that retains the essential
physical ingredients: the clusters gain and lose monomers by a surface reaction  that depends on the cluster size and the monomer concentration \emph{seen on the surface}.
Monomers outside the cluster undergo diffusion with finite diffusivity.
These ingredients give rise to a free boundary problem for the
growth of a cluster that contains  a new intrinsic cluster size,
$k_*$, in addition to the well known \emph{critical size}, $k_c$.
The critical size, $k_c$, separates shrinking clusters ($k<k_c$) from growing ones ($k>k_c$).
The new cluster size, $k_*$, indicates the importance of diffusion: 
the new prediction for cluster growth asymptotes to BD for small clusters with $k\ll k_*$, and to LS for large clusters with $k\gg k_*$.
In the former case, the diffusion effectively equates the surface density of monomer with the far-field density, thus, the surface reaction dictates the growth.
In the latter case, growth is strongly limited by finite diffusivity.
Furthermore, the  new model of cluster growth interpolates between BD and LS for intermediate cluster sizes on the order of $k_*$.

The smooth interpolation between BD and LS is crucial for a global model of aggregation that describes the whole process, from the initial creation of clusters from pure monomer, to the late stage growth-attrition process called \emph{coarsening}.
The essential idea is simple: if $k_c\ll k_*$, as expected in most cases, standard BD describes the \emph{nucleation} of super-critical ($k>k_c$) clusters and their growth while $k_c<k\ll k_*$. 
The super-critical clusters rapidly grow to sizes $k\gg k_*$, and their subsequent careers are described by LS.

Mathematically, we model this physics by a continuum approximation of
the discrete kinetics. 
The continuum equations constitute a PDE signaling problem for the distribution $r(k,t)$ of large ($k\ll k_*$) clusters in the space of (continuous) cluster size $k$.
At the lowest order of approximation, the cluster-size distribution satisfies an advection PDE, in which the growth rate ($\dot k$ vs. $k$) furnishes  the advection velocity.
The classical Zeldovich formula \cite{ZELD43}, which follows from BD, computes the creation rate  of super-critical ($k>k_c$) clusters.
Since we assume $k_c\ll k_*$, the Zeldovich formula gives rise to an effective source boundary condition on $k=0$.
The initial state of  pure monomers is expressed by a  zero initial condition, $r(k,0)\equiv 0$. 
Information about the amount of small ($k<k_c$) clusters is not expressed directly in $r(k,t)$. 
Instead, using conservation of particles, we express the amount of sub-critical clusters using an integral of $r(k,t)$.

Our theory does not handle nucleation that happens with $k_c$ on the order of $k_*$.
For this we suspect that a new theory is needed, one that considers the discrete and fluctuating nature of the monomer bath, and does not resort to
the diffusion equation, which arises from mean-field averaging.

The chapter is structured as follows.
In section \ref{sec:BD} we present a short summary of the classical microscopic
aggregation theory (BD). 
We derive rate constants for attachment and dissociation of monomers from a cluster by using free energy and detailed balance arguments.
The only difference from the classical theory is that it is based on
the \emph{surface} density of monomers, the density of monomers just outside the cluster, and does not assume that the monomer density is homogeneous.

In section 3 we take into account the finite spatial diffusion of
monomers. 
While still focusing on a single
cluster, we connect the surface monomer density with a far-field monomer
density, the nearly uniform concentration of monomers far from any cluster. 
This prescribes the growth rate of a cluster as a function of its size and the far-field monomer density. 
The standard assumption in diffusion limited aggregation is that the
surface density corresponds to a critical  cluster i.e.,
growing and shrinking are equally likely.
This seems paradoxical for two reasons. 
First, the free energy of a cluster as a function of cluster size, $k$, has its global maximum at $k=k_c$. 
On the face of it, this seems to be an unstable equilibrium, but yet it is claimed that the cluster remains at the top of this equilibrium.
In addition, if that is the value of the monomer density, how
does the cluster grow or shrink? 

This paradox is another artifact of assuming uniform monomer concentration as in BD\@.
It is deconstructed at the end of section 3, by an asymptotic analysis which exposes the stabilizing role of finite monomer diffusion:
If monomer concentration at the surface of a large ($k\gg k_*$) cluster has large deviation from the critical value described above, the surface reactions rapidly absorb or expel monomers.
Consequently, due to finite diffusivity, the surface monomer concentration undergoes a collateral adjustment \emph{towards} the critical value and the rapid reactions are turned off.
The growth rate due to diffusion is much slower and dictates the evolution of large clusters.

In section \ref{sec:zeldovich} we turn to the \emph{ensemble} of all
clusters.
If the density of monomers is below a certain \emph{saturation
  value},  an equilibrium exists, in which large clusters are extremely
unlikely.
For a `super-saturated' ensemble, with monomer density exceeding the
saturation value, clusters greater that a `critical cluster size',
$k_c$, have a strong tendency to persist and grow.
In this super-saturated case, there is no equilibrium; the
distribution of clusters is continuously changing.
Initially, there is \emph{nucleation}, which is the creation of
super-critical clusters. 
The calculation of the nucleation rate based on BD is reviewed here. 

As stated before, this chapter proposes a PDE signaling problem for the distribution of cluster sizes that quantifies the complete evolution of the aggregation process.
Section 5 contains the assembly of the signaling problem from the component
parts in sections 2--4.
It has a peculiar nonlinearity, in which the advection velocity in the PDE and the boundary condition (BC) at $k=0$ depend on the monomer
density as a parameter.
The monomer density can be written as an integral of the solution, and herein lies the nonlinearity.

The nonlinearity makes the task of solving the equations difficult
enough to warrant placing it in the next chapter.
\section{Classical Becker-D\"oring Model}
\label{sec:BD}
Becker-D\"oring theory (BD) imposes simplifying assumptions at the outset:
The clusters are assumed to be uniformly distributed in a dilute
`bath' of monomers. 
This assumption is adjusted regarding the distribution of
monomers in the next section, when we add diffusion. 
The clusters are assumed to change size only by losing or gaining
one monomer at a time. 
Two large clusters do not fuse together nor does one cluster
 break into two. 
This can be justified heuristically by noticing that the density of
the large clusters is much smaller than the (already small) density of
monomers, thus the probability of two large clusters interacting is
small. 
In addition, the mobility of the large clusters is much smaller than that of the monomers, so they are even less likely to stumble upon one another. 
Similarly, since the large clusters have a low mobility (relative to monomers) a cluster that breaks into two will, most likely, reconnect quickly, as the two parts remain close together. 

Another important assumption is that the only governing parameter of a cluster is its size. 
The shape of the cluster is assumed to be fixed. 
This assumption can be weakened to require that clusters of same size have the same binding energy and the same surface area.

To derive the kinetic model of nucleation we introduce the essential quantitative ingredients: energy,
free energy, and the rate constants of transitions between configurations.
\subsection{Energy and Free Energy}
%
%

\begin{figure}[ht!]
\centerline{ \parbox{8cm}{\centerline{
\resizebox{6cm}{!}{\includegraphics{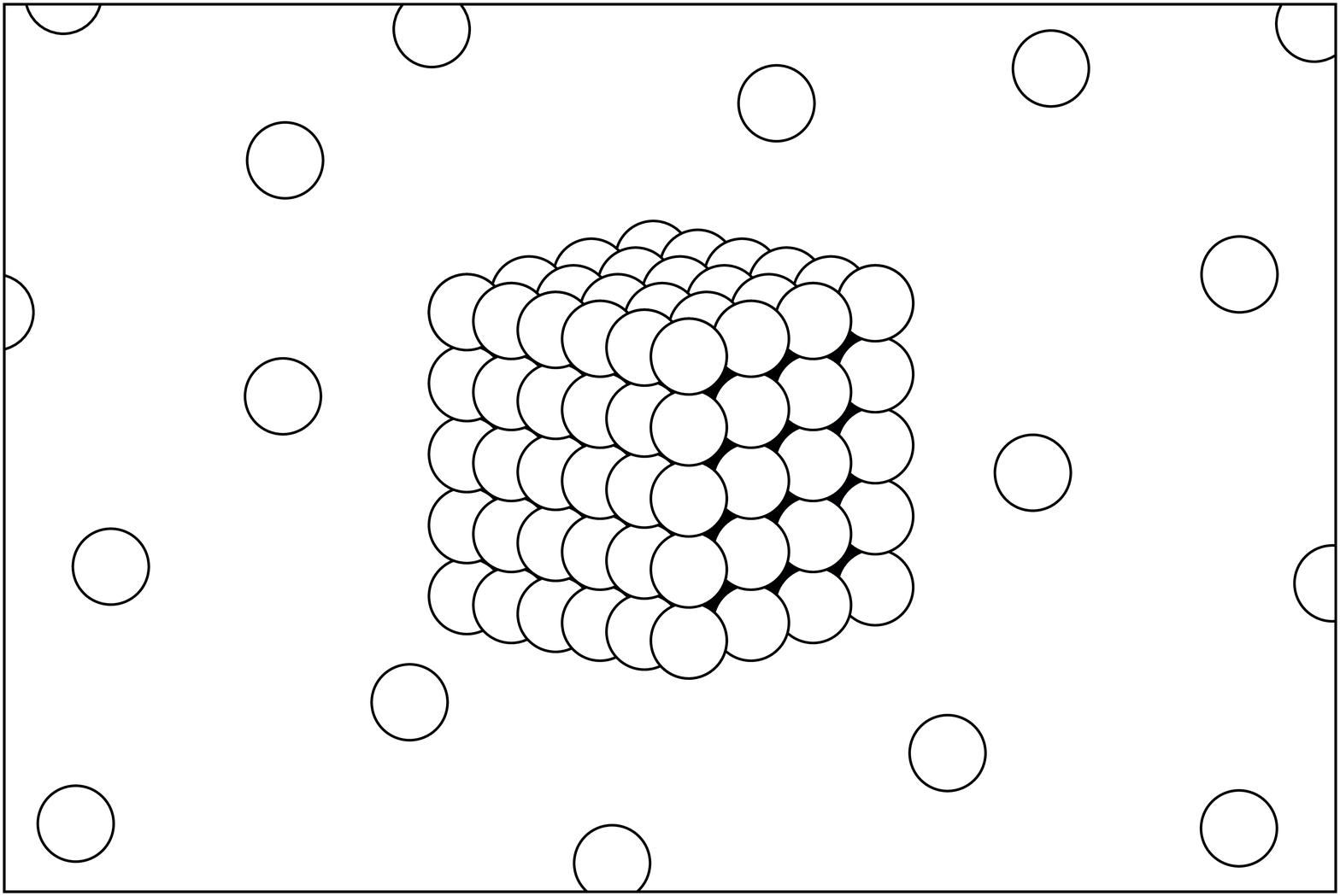}}}
\caption[Schematic cubic crystal]{A schematic cubic crystal surrounded by the monomer bath.}}}
\end{figure}
The energy of cubic cluster with $k$ monomers is
\begin{equation}
  \eps_k \approx 3 \eps (k-k^\frac23).\label{eq:energy:cube}
\end{equation}
Here $\eps$ is the binding energy of a single bond between two
adjacent particles, the energy needed to break it. 

While the clusters are not assumed to be simple cubes,  the general structure of the binding energy is expected to remain. 
Thus, we assume that there is a bulk energy constant, $\alpha>0$, and a surface energy constant, $\sig>0$, such that the energy of a $k$-cluster is
\begin{equation}
  \eps_k \approx k_BT(\alpha k-\frac32\sig k^\frac23), \text{ for } k\gg1, \label{eq:binding:energy}
\end{equation}
where, $k_BT$ is the Boltzmann factor.
The factor of $\frac32$ is added in hind-foresight, as it makes some formulas cleaner.
Equation \eqref{eq:binding:energy} is only true asymptotically for large clusters. 
For small clusters, the separation between `bulk' and `surface' is
artificial, and we do not expect \eqref{eq:binding:energy} to be
quantitatively accurate. 
In particular, $\eps_1=0$ since the binding energy is the change in the
energy from the unbound state and a cluster with one particle is
unbound.

Next we consider the \emph{free energy} costs to create a $k$-cluster from the monomer bath.
The bath is characterized by the density $\rho_1$ of monomer, measured in units of $\oneover{v}$, where $v$ is the volume of each monomer.
In other words, $\rho_1$  is the volume fraction occupied by monomers

The free energy cost to create a $k$-cluster from the monomer bath is 
\begin{align}
  g_k &= -\eps_k -k_BT k \log \rho_1\notag\\
&\approx -k_BT\brk{(\alpha+\log \rho_1)k-\frac32\sig k^\frac23},
  \text{ for } k\gg 1.\label{eq:free:energy}
\intertext{
Here, $k_B T \log \rho_1$ is the chemical potential of a monomer in
the bath.
Rewriting the $k\gg1$ asymptotic form of the free energy gives insight
into the existence of a critical monomer density, $\rho_s \equiv
e^{-\alpha}$. 
We call $\rho_s$ the \emph{saturation density} of monomers.
Setting $\alpha=\log \oneover{\rho_s}$ in \eqref{eq:free:energy} gives}
  g_k&\approx k_BT\brk{\frac32\sig k^\frac23-k \log \frac{\rho_1}{\rho_s}}.\notag
\end{align}
Thus, when $\rho_1<\rho_s$ the free energy increases with $k$, allowing for an equilibrium. 
When $\rho_1>\rho_s$, the free energy attains its maximum at the critical cluster size $k_c$,

\begin{equation}
  k_c\approx \brk{\frac{\sig}{\log \frac{\rho_1}{\rho_s}}}^3, \text{
  as } \rho_1\goto\rho_s^+.\label{eq:critical:k}
\end{equation}
We  investigate the implications of this critical value later.
\subsection{Kinetics and Detailed Balance}
%
%

\begin{figure}[h!]
  \centerline{
\parbox{8cm}{
\centerline{\resizebox{6cm}{!}{\includegraphics{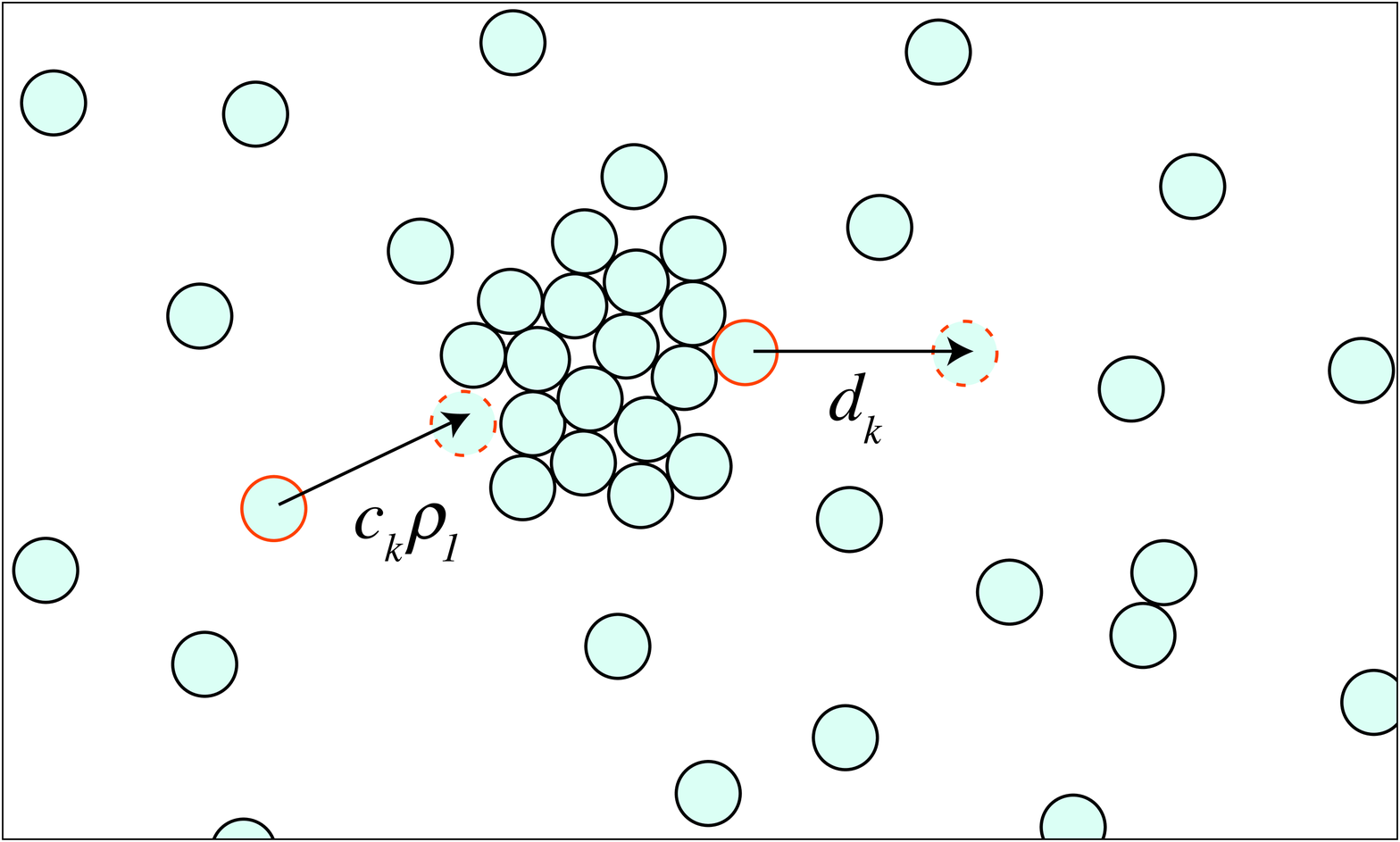}}}
\caption[Surface activated growth processes]{The exchange of particles between a cluster and the monomer
  bath. $\rho_1c_k$ is the rate at which a monomer gets added to the
  cluster, and $d_k$ is the rate at which monomers leave.}}}
\end{figure}

To study the rate of change in the cluster size we need to model the allowed reactions. 
BD allows two types of reactions:(1) A monomer in the bath can join the cluster; (2) A particle on the surface of the cluster can  dissociate from it and enter the bath. 
Mathematically, the two reactions are modeled as independent Poisson processes.
Let us call the rate at which particles leave a $(k+1)$-cluster $d_k$, and
the rate at which particles join a $k$-cluster $c_k\rho_1$.
We work under the assumption that the solution is dilute, that is,
$\rho_1\ll 1$.
It is reasonable to expect that in the dilute limit, the growth
rate is proportional to $\rho_1$.
Thus, we include $\rho_1$ in the growth rate, $c_k\rho_1$, so that both $c_k$ and
$d_k$ are independent of $\rho_1$. 

Initially we assume a uniform $\rho_1$, so it is a global parameter. 
When we add diffusion, we are a little more careful and take $\rho_1$
to be the density at the surface of the cluster. 

Since the dissociation happens on the boundary of the cluster, we
expect $d_k$ to be proportional to the surface area, which, in turn,
is proportional to $k^{2/3}$,
\begin{equation}
  d_k = \omega k^\frac23.\label{eq:d_k}
\end{equation}
Here, $\omega$ includes a per-particle  dissociation rate and a geometric factor.

A standard \emph{detailed balance} argument relates $c_k$ to $d_k$.
Let $\rho_1$ be the value of monomer density so that a $k$-cluster is in
equilibrium with the monomer bath. 
That is, the cluster has no net tendency to grow nor shrink, hence, the adsorption rate, $\rho_1c_k$, should match the emission rate, $d_k$.
The physical requirement for the equilibrium is that the free
energy is unchanged when a monomer is taken from the bath and added to the cluster,
\begin{equation*}
  g_{k+1}-g_k=-\eps_{k+1} + \eps_k -k_BT\log\rho_1 =0,
\end{equation*}
hence,
\begin{equation*}
  \rho_1 = e^{\frac{\eps_{k} - \eps_{k+1}}{k_BT}}.
\end{equation*}
The detailed balance relation between $c_k$ and $d_k$ is therefore,
\begin{equation}
  c_k = e^{\frac{\eps_{k+1} - \eps_{k}}{k_BT}}d_k. \label{eq:detailed:balance}
\end{equation}
We  use the adsorption and emission rates, $c_k\rho_1$ and $d_k$, to
derive a kinetic equation for the expected change in size of a cluster. 
In a small time span $\delta t$, the expected  change in cluster size,
$\delta k$, is:
\begin{equation}
  \Average{\delta k} = ( c_k\rho_1 - d_k) \delta t.\label{eq:expected:size:change}
\end{equation}
Using the model for the binding energy of large clusters \eqref{eq:binding:energy}
and  relation \eqref{eq:detailed:balance} between $c_k$ and $d_k$,
equation \eqref{eq:expected:size:change} can be rewritten as
\begin{equation*}
  \Average{\delta k} \approx \omega \brk{
 \frac{\rho_1-\rho_s}{\rho_s} k^\frac23-\sig k^\frac13 } \delta t, \text{ for } k\gg 1.
\end{equation*}
In the appendix we show that
if the variance in $k$ is much smaller than $\Average{k}$ initially,
it will remain so, as long as $\Average{k}$ is bounded away from the
critical size $k_c$ in \eqref{eq:critical:k}.
In this case we approximate the evolution of $k(t)$ for a given
cluster as deterministic, and governed by the ODE
\begin{equation}
  \dot{k} = \omega \brk{\eta k^\frac23 - \sig k^\frac13 }.\label{eq:BD:kinetics}
\end{equation}
Here, $\eta$ is the \emph{super-saturation}, defined by
\begin{equation}
  \eta = \frac{\rho_1-\rho_s}{\rho_s}.\label{eq:supersaturation}
\end{equation}
The super-saturation in \eqref{eq:supersaturation} is generally a function of time, due to the exchange of particles between clusters and the monomer bath.
By conservation of the \emph{total} particle density, the average value
of $\rho_1$, and hence $\eta$, is determined from the densities of all
clusters with $k\ge2$. 
Within the framework of BD, which assumes that the monomer density is uniform, we simply
set $\eta$ to this average value, and, in this sense,
\eqref{eq:BD:kinetics} is the BD prediction for cluster growth.
However, if the diffusivity of monomers is finite, the density of
monomers seen at the surface of the cluster will be different from the
average value far away.
We propose that \eqref{eq:BD:kinetics} holds generally, with $\eta$
equal to the value of super-saturation seen at the \emph{surface} of
the cluster.
Equations \eqref{eq:BD:kinetics} and \eqref{eq:supersaturation} expose
the \emph{kinetic} significance of the saturation density
$\rho_s= e^{-\alpha}$, and the critical cluster size $k_c$ in
\eqref{eq:critical:k}, which we rewrite using $\eta$,
\begin{equation}
  k_c\approx
  \brk{\frac{\sig}{\eta}}^3, \text{ as } \eta\goto 0^+.\label{eq:critical:k:2}
\end{equation}
If the surface value of $\rho_1$ is less than $\rho_s$, all the clusters
shrink, regardless of $k$.
For $\rho_1>\rho_s$, i.e. $\eta>0$, the critical size, $k_c$, separates growing,
\emph{super-critical} clusters ($k>k_c$) from shrinking, \emph{sub-critical} ones ($k<k_c$).

It should be noted that while the expected change in size of sub-critical clusters is negative, there is a small probability for a sub-critical cluster to grow and become super-critical.
The rate at which this happens is estimated by the Zeldovich
formula, to which  we come back in section \ref{sec:zeldovich}.
First we add finite diffusion of monomers to the model and see how it
affects the growth rate.


\begin{figure}[h!]
  \centerline{
\parbox{8cm} {\centerline{
\resizebox{6cm}{!}{\includegraphics{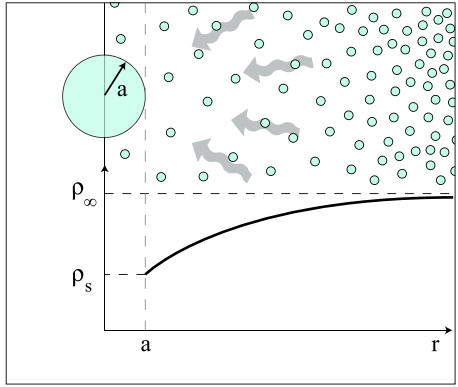}}}
\caption[Diffusion driven growth]{The cluster is surrounded by a \emph{inhomogeneous} monomer
  bath. The flux of monomers in the bath is diffusive, and the cluster
  reacts to the \emph{local} monomer density.}
}}
\end{figure}

\section{Adding Monomer Diffusion}
Diffusion is the usual model of transport for monomers in the bath.
For a finite diffusion coefficient, $D$, the exchange of particles
between a cluster and the bath directly outside of it leads to
non-uniform monomer density.
Hence, the monomer concentration, $\rho_1$, is a function of position and 
time that satisfies the diffusion PDE.

To add diffusion to the BD model, we make a couple of additional
assumptions. 
The cluster is taken to be spherical, filled with monomers, each
taking volume $\nu$.
Thus, the number of particles in the cluster, $k$, and the radius of
the cluster, $a$, are related by
\begin{equation}
  k\nu = \frac{4\pi}{3}a^3.\label{eq:cluster:radius}
\end{equation}
The monomer density is assumed to be radially symmetric (with the
cluster centered at the origin.) 
Thus, the density, $\rho_1(r,t)$, satisfies the radially symmetric diffusion PDE in $\R^3$
\begin{align}
   \partial_t \rho_1 &= \frac{D}{r^2} \partial_{r}(r^2 \partial_r \rho_1) & &\text{ in } r>a,\label{eq:diffusion}
\end{align}

There are two BC at $r=a$: one is the kinetics due to BD as in equation
\eqref{eq:BD:kinetics}.
That is, $\rho_1(a,t)$ is related to $k$ and $\dot k$ by
\begin{equation}
  \dot{k} = \omega \brk{\eta(a,t) k^\frac23 - \sig k^\frac13 }\label{eq:BD:kinetics:2},
\end{equation}
where, 
\begin{equation*}
  \eta(a,t) = \frac{\rho_1(a,t)-\rho_s}{\rho_s}
\end{equation*}
is the super-saturation seen at the surface of the cluster.

The second BC results from conservation of particles.
The particles that are added to the cluster can come from two possible
sources: particles in the solution surrounding the cluster, which join
the cluster as it engulfs them, and  particles added by the diffusive flux.
This can be put in a simple equation:
\begin{equation}
  \dot a ( 1-\rho_1(a,t)) = D (\partial_r \rho_1)(a,t)=D\rho_s(\partial_r\eta)(a,t).\label{eq:FBP}
\end{equation}

Finally, there is a BC at $\infty$: 
the monomer concentration has the asymptotically uniform value
$\rho_\infty$ far from the clusters,
\begin{equation}
    \rho_1\goto\rho_\infty \text{ as } r\goto\infty.\label{eq:far:field}
\end{equation}
In the context of the full aggregation problem, $\rho_\infty$ is
generally a function of time, which  follows from overall conservation
of particles.
Equations (\ref{eq:cluster:radius}--\ref{eq:far:field}) constitute a
free boundary problem (FBP) for $a(t)$ and $\rho_1(r,t)$ in $r>a$.

The analysis of the FBP begins by identifying suitable non-dimensional
variables.
We assume that the local super-saturation
\begin{equation*}
  \eta(r,t)\equiv \frac{\rho_1(r,t)-\rho_s}{\rho_s}.\label{eq:super:saturation}
\end{equation*}
is uniformly small in $r>a$.
Hence, we introduce $\eps$ as a gauge parameter for $\eta(r,t)$ and we
replace $\eta(r,t)$ in (\ref{eq:cluster:radius}--\ref{eq:far:field})
by $\eps\eta(r,t)$.
The analysis of the FBP is based on an $\eps\goto0$ limit process.
However, we do not yet take the $\eps\goto0$ limit. 
It remains to determine the scaling of the other variables $k,r,a$ and
$t$ with $\eps$.
These follow from simple physical balances: 
in the `kinetic' BC \eqref{eq:BD:kinetics:2}, $\eta(r,t)$ has order of
magnitude $\eps$, so the two terms of the RHS balance when $k$ has
magnitude $\brk{\frac{\sig}{\eps}}^3.$
Notice that this is the critical cluster size \eqref{eq:critical:k:2}
for $\eta=\eps$.
The scalings of $\eta$ and $k$ are recorded in the scaling table
\begin{center}
Scaling Table\\
\begin{tabular}{l*{4}{>{$}c<{$}}}
Variable&\eta&k&a,r&t\\
\hline\\[-4mm]
Unit&\eps&\brk{\frac{\sig}{\eps}}^3&\frac{\sig}{\eps}\nu^{1/3}&\frac{\sig}{\omega\eps^2}
\end{tabular}
\end{center}
The unit of $r$ and $a$ is the radius of a cluster with
$k=\brk{\frac{\sig}{\eps}}^3$ particles.
The unit of $t$ is chosen to balance the LHS of the BC
\eqref{eq:BD:kinetics:2} with the two terms on the RHS.

The equations of the non-dimensional FBP that follow from
(\ref{eq:cluster:radius}--\ref{eq:far:field}) are 
\begin{align}
  k&=\frac{4\pi}{3}a^3,\label{eq:cluster:radius:ND}\\
  \partial_t \eta&=\frac{D}{\omega \nu^{2/3}\sig} \oneover{r^2}\partial_{r}(r^2\partial_r\eta), \text{ in } r>a, \label{eq:diffusion:ND} \\
    \dot k &= \eta(a,t)  k^{2/3} -k^{1/3},\label{eq:BD:scaled}\\
  \dot a \brk{\rule{0cm}{4mm}1-\rho_s(1+\eps\eta(a,t))} & = \frac{D}{\omega \nu^{2/3}}
  \frac{\eps\rho_s}{\sig}  (\partial_r\eta)(a,t)\label{eq:conservation:ND}\\
\eta&\goto\eta_\infty \text{ as } r\goto\infty. \label{eq:eta:infty}
\end{align}
Here, $\eta_\infty$ is the asymptotically uniform value of $\eta(r,t)$
as $r\goto\infty$.
The characteristic time of its variation is assumed to be comparable
to or larger than the unit of time $\frac{\sig}{\omega\eps^2}$ from the
scaling table.

The dimensionless constant $\frac{D}{\omega \nu^{2/3}}$ can be interpreted as a ratio of characteristic times for two different physical processes.
Recall that $\oneover{\omega}$ is the characteristic time for a monomer on the surface of a cluster to dissociate into the solution. 
The ratio $\frac{\nu^{2/3}}{D}$ is the characteristic time for a monomer to diffuse a distance comparable to its own size.
The conventional assumption is that the ``dissociation time'' is much longer than the ``diffusion time,'' so that 
\begin{equation*}
  \oneover{\omega}\gg \frac{\nu^{2/3}}{D} \quad \Longleftrightarrow \quad \frac{D}{\omega \nu^{2/3}}\gg1. 
\end{equation*}
In this limit, the diffusion equation \eqref{eq:diffusion:ND} reduces to a radial Laplace equation
\begin{equation*}
  \partial_{r}(r^2\partial_r\eta)=0.
\end{equation*}
The solutions with  $\eta=\eta_\infty$ as $r\goto\infty$ are
\begin{equation}
  \eta(r,t) = \eta_\infty + (\eta(a,t)-\eta_\infty)\frac ar.\label{eq:eta:lap}
\end{equation}
The time dependence of $\eta$ is implicit due to the time-dependence
of its values at $r=a$ and $r=\infty$.

In the dilute limit, with $\rho_s\ll1$, the conservation equation \eqref{eq:conservation:ND} reduces to
\begin{equation*}
  \dot a = \frac{D}{\omega \nu^{2/3}} \frac{\eps \rho_s}{\sig} (\partial_r \eta)(a,t).
\end{equation*}
Substituting $\eta$ from \eqref{eq:eta:lap}, this is becomes
 \begin{equation}
  a \dot a = \frac{D}{\omega \nu^{2/3}} \frac{\eps \rho_s}{\sig} \brk{\eta_\infty-\eta(a,t)\rule{0mm}{4mm}}.\label{eq:a:dot:diff}
\end{equation}
Equation \eqref{eq:a:dot:diff} can be converted into an equation for
$\dot k$ using \eqref{eq:cluster:radius:ND}. 
This gives
\begin{equation}
  \dot k = \eps\mu(\eta_\infty-\eta(a,t))k^\third, \label{eq:k:dot:diff}
\end{equation}
where $\mu$ is the dimensionless constant:
\begin{equation*}
  \mu = (3 \cdot 16 \pi^2)^\third \brk{\frac{D}{\omega\nu^{2/3}}}\brk{ \frac{\rho_s}{\sig}}.
\end{equation*}
Since $\mu$ is a product of a large number, $\frac{D}{\omega \nu^{2/3}}$,
and a small number, $\frac{\rho_s}{\sig}$, it can be large or small. 
We therefore entertain any value of $\mu$.

The two equations for $\dot k$, \eqref{eq:BD:scaled} and
\eqref{eq:k:dot:diff},  involve the super-saturation at the surface of the cluster, $\eta(a,t)$. 
Solving for  $\dot k$ and $\eta(a,t)$ gives
\begin{align}
  \dot k &= \frac{\eta_\infty k^{2/3}-k^{1/3}}{1+\frac{k^{1/3}}{\eps\mu}}, \label{eq:new:cluster:kinetics}\\
  \eta(a,t) &= \frac{\eta_\infty + \oneover{\eps\mu}}{1 + \frac{k^{1/3}}{\eps\mu}}.\label{eq:eta_s}
\end{align}
ODE \eqref{eq:new:cluster:kinetics} indicates a \emph{second} characteristic cluster size besides $k_c$. 
In units of $\brk{\frac{\sig}{\eps}}^3$ this cluster size is
  \begin{equation*}
    k_* = (\eps\mu)^3.
  \end{equation*}
In the original variables, $k_*$ is a combination of basic physical constants:
\begin{equation}
  k_* = (\sig\mu)^3=(3 \cdot 16 \pi^2) \frac{D^3\rho_s^3}{\omega^3 \nu^2}.\label{eq:k_*}
\end{equation}
Notice that for $k\ll k_*$  equation \eqref{eq:new:cluster:kinetics}
asymptotes to
BD\@.
For $k\gg k_*$, the asymptotic form of
\eqref{eq:new:cluster:kinetics} is
\begin{equation}
  \dot k \approx \eps\mu\brk{\eta_\infty k^\third-1}.\label{eq:LS}
\end{equation}
Restoring original units, \eqref{eq:LS} becomes 
\begin{equation}
  \dot k = d \brk{\eta_\infty k^\third-\sig},\quad d = (3\cdot
  16\pi^2)^\third\frac{D\rho_s}{\nu^\frac23}, \label{eq:LS:dim}
\end{equation}
which is the standard result for diffusion limited growth (DLG)
\cite{LS61}.
Equation \eqref{eq:eta_s} shows how the surface value of super-saturation differs from the uniform value, $\eta_\infty$, far from the cluster. 
Notice that it is a function of $k$. 
We convert it into an equation for $s_k$, the value of monomer density
seen at the surface of a $k$-cluster (again, in original units):
\begin{equation}
  s_k=\rho_s\brk{1+\frac{\eta_\infty+\oneover{\mu}}{1+\frac{k^{1/3}}{\sig\mu}}}.\label{eq:s_k}
\end{equation}
This will be important, when we examine the whole ensemble of clusters, and formulate evolution equations for cluster densities.
\subsection{Physical Meaning of $k_*$}
We show that $k_*$ is the characteristic size of clusters for which finite diffusion induces a significant relative difference between $\eta_\infty$ and $\eta(a,t)$.
That is, $\delta\eta\equiv\eta_\infty-\eta(a,t)$ is comparable in magnitude to $\eta_\infty$.
A simple examination of two physical balances is sufficient. 
First, the cluster's growth rate balances the diffusive influx of monomers.
This is expressed by
\begin{equation}
  \dot k = a^2 D \frac{\brk{\frac{\rho_s}{\nu}}\delta\eta}{a}.\label{eq:diffusion:mag}
\end{equation}
Here, the equality means `order of magnitude balance'. 
In the RHS, $\frac{\rho_s}{\nu}\delta\eta$ is the difference between monomer densities at $\infty$, and on the surface, expressed in the conventional unit of 1/volume. 
For quasi-static diffusion, the diffusion zone about the cluster of radius $a$ has thickness $a$, so $\frac{\brk{\frac{\rho_s}{\nu}}\delta\eta}{a}$ estimates $(\partial_r\rho_1)(a,t)$ and the influx of monomers per unit area into cluster is estimated by multiplying this by $D$. 
Finally, multiplying by the area, proportional to $a^2$, gives the cluster growth rate, $\dot k$. 

Second, the magnitude of $\dot k$ as dictated by the surface reactions \eqref{eq:BD:kinetics:2} is 
\begin{equation}
  \dot k = \omega\eta k^\frac23.\label{eq:surface:reaction:mag}
\end{equation}
Enforcing the equivalence of \eqref{eq:diffusion:mag} and
\eqref{eq:surface:reaction:mag} and using $k=\nu a^3$ (order of magnitude equality again), we find
\begin{equation*}
  \frac{\delta\eta}{\eta}=\frac{\omega \nu^{2/3}}{D\rho_s}k^\third.
\end{equation*}
And we see that $\delta\eta$ is comparable to $\eta$ when $k$ is comparable to $k_*=(\sig\mu)^3$.

\subsection{Critique of DLG and its `paradox'}
We briefly examine the `traditional' derivation of ODE \eqref{eq:LS} for DLG, within the framework of the non-dimensional free boundary problem (\ref{eq:cluster:radius:ND}--\ref{eq:eta:infty}).
Given $\eta(a,t)$, equation \eqref{eq:k:dot:diff} gives the growth rate
of the cluster that follows from diffusive flux of monomers.
In the traditional analysis of DLG, $\eta(a,t)$ is chosen so that the
cluster is in equilibrium with the monomer bath that surrounds it.
Under the current non-dimensionalization, this `critical nucleus' BC reads
\begin{equation}
  \eta(a,t) = k^{-\third}\label{eq:critical:ND}.
\end{equation}
Substituting \eqref{eq:critical:ND} for $\eta(a,t)$ in
\eqref{eq:k:dot:diff} leads directly to the ODE \eqref{eq:LS}.

By inspection, we see that the traditional equations
(\ref{eq:LS}, \ref{eq:critical:ND}) arise by taking the $\eps\mu\goto0$
limit of (\ref{eq:new:cluster:kinetics}, \ref{eq:eta_s})  with $k$
fixed.
The alternative limit process, $\frac{k}{k_*}\goto\infty$ with
$k_*=\eps\mu$ fixed is more physical: the value of $\eps\mu$ is set by
material properties and initial conditions, and we expect that
clusters eventually grow to sizes $k\gg k_*$. 
We have already seen that the ODE \eqref{eq:new:cluster:kinetics} for
$k$ converges to the DLG result in this limit, but the expression
\eqref{eq:eta_s} for the surface value of monomer density does not
converge to the DLG boundary condition \eqref{eq:critical:ND}.
Instead, 
\begin{equation}
  \eta(a,t) \approx (1+\mu \eta_\infty)k^{-\third},\label{eq:surface:mono:den}
\end{equation}
which has an additive term $\mu\eta_\infty$ in the prefactor of
$k^{-1/3}$ not present in $\eqref{eq:critical:ND}$.
A mathematical critique of the `critical nucleus' boundary condition
\eqref{eq:critical:ND} is simple; it results from formally neglecting
$\dot k$ in the LHS of the `surface kinetics' boundary condition
\eqref{eq:BD:scaled}.
In our result, $\dot k$ balances the RHS, even in the limit
$\frac{k}{k_*}\goto\infty$, resulting in \eqref{eq:surface:mono:den}.

Recall that the `critical nucleus' boundary condition in
traditional DLG looks paradoxical because the `cluster sits on top of
a free energy maximum'.
A `lazy' deconstruction might say: ``Nothing to explain, the critical
nucleus boundary condition is simply incorrect in the (more physical)
limit $\frac{k}{k_*}\goto\infty$ with $k_*$ fixed.'' 
Another easy explanation  looks at the free energy. The free energy
\eqref{eq:free:energy} refers to a simple cluster surrounded monomers of
\emph{uniform}  density, whereas the actual kinetics we consider
involves a \emph{non-uniform} density $\rho_1(r,t)$ in $r>a$.
The actual free energy takes into account the functional dependence of
$\rho_1(r,t)$ in $r>a$.
These remarks indicate that the `paradox' in its original form is
na\"ive.
Nevertheless, it points to some physics that is not expressed in the
quasi-static model (\ref{eq:new:cluster:kinetics}, \ref{eq:eta_s}) of cluster growth as it stands.

Suppose that we place a cluster of size $k$ into a uniform monomer bath
that has the `wrong` monomer density, not equal to the surface value
$s_k$ given in \eqref{eq:s_k}.
In order for our model to be
plausible, the surface value of $\rho_1$ should rapidly relax to $s_k$
in  \eqref{eq:s_k}. 
We now show that the full free boundary problem
(\ref{eq:cluster:radius:ND}--\ref{eq:eta:infty}) implies such a
relaxation transient.

\subsection{The Stability of the Free Boundary Problem}
The relaxation transient is characterized by a balance of the time and
space derivatives in the diffusion PDE \eqref{eq:diffusion}.
Hence the characteristic time of the relaxation transient is
\begin{equation}
  t_r \equiv \frac{a^2}{D},\label{eq:relax:time}
\end{equation}
where $a$ is the cluster radius.
The relative change of the cluster radius in this characteristic time
is small: from \eqref{eq:FBP}, $\dot a$ has the order of magnitude
$\frac{D\rho_s\eps}{a}$.
Hence, the relative change of radius in time $t_r$ has magnitude
$\eps\rho_s$.
The small relative change in cluster radius means that the cluster
radius is asymptotically constant during the relaxation transient, and
it remains to derive from the full free boundary problem
(\ref{eq:cluster:radius}--\ref{eq:far:field}) a reduced boundary value
problem for $\eta(r,t)$ in $r>a$, with $a$ fixed.
We use the previous units in the scaling table for all variables
except time $t$. 
For $t$ we use $t_r$ in \eqref{eq:relax:time} with 
$a$ replaced by the characteristic cluster radius,
$\frac{\sig}{\eps}\nu^{1/3}$.
The reduced boundary value problem is
\begin{align}
\partial_t \eta = \oneover{r^2} \partial_r(r^2 \partial_r \eta)\text{
  in } r>a,\label{eq:trans:PDE}\\
\lambda(\partial_r\eta)(a,t) = \eta(a,t) -
  k^{-\third},\label{eq:trans:BC}\\
\eta(r,t)\goto\eta_\infty\text{ as } r\goto\infty,\label{eq:trans:infty}
\end{align}
in the limit $\eps\goto0,$ and $\lambda \equiv\brk{\frac{3}{4\pi}}^\third
  \eps \mu$ fixed.
The far-field super-saturation, $\eta_\infty$, is assumed to vary on a
characteristic time much longer than $t_r$, so $\eta_\infty$ is
effectively constant.

The time-independent solution of
(\ref{eq:trans:PDE}--\ref{eq:trans:infty}) for $\eta(r,t)$ is
\eqref{eq:eta:lap} with $\eta$ on $r=a$ given by \eqref{eq:eta_s}.
We show that this time-independent solution is asymptotically stable.
We notice that
\begin{equation}
  E\equiv\frac{\lambda}{2}\int_a^\infty r^2(\partial_r\eta)^2  \, dr +
  \frac{a^2}{2} \brk{\eta(a,t)-k^{-\third}}^2\label{eq:Lyapunov}
\end{equation}
is a Lyapunov functional for equations
(\ref{eq:trans:PDE}--\ref{eq:trans:infty}).
The time derivative of $E$,
\begin{equation*}
  \dot E = -\lambda \int_a^\infty r^2\eta_t^2 \,dr,\label{eq:Lyapunov:deriv}
\end{equation*}
is found by time-differentiation of \eqref{eq:Lyapunov}, integration by parts, and use
of the PDE \eqref{eq:trans:PDE} and BC \eqref{eq:trans:BC}.
Since $E$ is positive definite, and $\dot E\le0$, it follows that $\eta(r,t)$ converges
to the time-independent solution. 
We conclude that if the surface monomer concentration is initially
different from the quasi-static value \eqref{eq:eta_s}, it relaxes to
it in characteristic time $t_r$.

\section{Evolving Distribution of Cluster Sizes}
\label{sec:zeldovich}
The kinetics equation \eqref{eq:new:cluster:kinetics} requires $\eta_\infty$, the super-saturation far from any cluster.
To find $\eta_\infty$ we look at the joint evolution of all the clusters, each assumed to follow the dynamics in \eqref{eq:new:cluster:kinetics}.
The clusters are coupled by the combined effect they have on the
monomer density, and consequently, on the super-saturation.

Let $\rho_k(t)$ be the average spatial density (in units of $\oneover{\nu}$) of $k-$clusters at time $t$. 
Assuming that the total particle density has a fixed value, $\rho$,
the $\rho_k$  satisfy a particle conservation equation. 
Since a $k$-cluster is made of $k$ particles, the total particle density, $\rho$, must satisfy
\begin{equation}
  \rho = \sum_{k=1}^\infty k \rho_k.\label{eq:conservation}
\end{equation}
With the help of \eqref{eq:conservation}, the space averaged super-saturation can be rewritten as a function of the cluster densities $\rho_k$ with $k\ge2$:
\begin{equation}
  \frac{\rho_1-\rho_s}{\rho_s} = \frac{\rho-\rho_s}{\rho_s} -\oneover{\rho_s}\sum_{k=2}^\infty k\rho_k.\label{eq:rho_1:con}
\end{equation}
In the dilute limit with inter-cluster distances much greater than cluster radii, we expect that the super-saturation is asymptotically uniform, throughout most of the monomer bath far from clusters.
In this case, that asymptotically uniform value, $\eta_\infty(t)$, should be well approximated by the spatial average \eqref{eq:rho_1:con},
\begin{equation}
  \eta_\infty=\frac{\rho-\rho_s}{\rho_s} -\oneover{\rho_s}\sum_{k=2}^\infty k\rho_k.\label{eq:super-sat:con}
\end{equation}

We turn to the  evolution of the densities.
The $\rho_k$ obey kinetic equations associated with the
reactions
\begin{equation*}
  k \text{-cluster} + \text{monomer} \rightleftarrows (k+1)\text{-cluster}.
\end{equation*}
The equations are
\begin{align}
  \dot{\rho_k}&=j_{k-1}-j_k,\label{eq:discrete:kinetic}
\intertext{for $k\ge2$, where the discrete flux $j_k$ is  the \emph{net} rate of creation of a $(k+1)$-cluster from a
$k$-cluster,}
j_k&\equiv c_ks_k\rho_k -d_k\rho_{k+1}.\label{eq:discrete:flux}
\end{align}
As before, $d_k$ is the rate constant for shedding a monomer from the
surface of a $(k+1)$-cluster.
For $k\gg1$ it has the asymptotic behavior \eqref{eq:d_k}, proportional to surface area.
The prefactor $c_ks_k$ of $\rho_k$ in \eqref{eq:discrete:flux} is the rate constant for \emph{adding} a monomer.
Recall that $s_k$ is the value of monomer density \emph{seen at the surface} of a $k$-cluster, given by \eqref{eq:s_k}, and $c_k$ is related to $d_k$ by the detailed balance condition \eqref{eq:detailed:balance}.
A more explicit formula for $j_k$ displaying the $k$-dependence of surface monomer concentration and detailed balance is
\begin{equation}
j_k =   d_k\brk{e^{\frac{\eps_{k+1}-\eps_k}{k_BT}}s_k\rho_k-\rho_{k+1}}.\label{eq:j_k}
\end{equation}
Equations (\ref{eq:discrete:kinetic}, \ref{eq:j_k}) can be summarized as discrete advection-diffusion equations,
\begin{equation}
  \dot\rho_k+D^-\Brk{d_k\brk{1-e^{\frac{\eps_{k+1}-\eps_{k}}{k_BT}}s_k}\rho_k-d_kD^+\rho_k}=0 \text{ for } k\ge2.\label{eq:discrete:ad:dif}
\end{equation}
Here $D^+, D^-$ are, respectively, the forward and backward difference operators.
In these equations, the surface monomer density $s_k$ contains the super-saturation $\eta_\infty$ as a parameter, and $\eta_\infty$ is connected to the $\rho_k$ according to \eqref{eq:super-sat:con}.
So we see explicitly how the cluster densities are coupled to each other via the super-saturation.
The $k$-dependence of $s_k$ induced by the finite diffusivity of monomers is the essential difference from classical BD\@. 
We recover classical BD by taking $\mu\goto\infty$, which in turn results from $\frac{D}{\omega \nu^{2/3}}\goto\infty$. 
Then $s_k$ in \eqref{eq:s_k} reduces to $\rho_s(1+\eta_\infty)$, which is the \emph{uniform} value of $\rho_1$ assumed in classical BD.

\subsection{Equilibrium}
\emph{Equilibria} are time independent densities, $\tilde\rho_k$, so that all the fluxes $j_k$ are zero, and the sum \eqref{eq:conservation}, which gives the total particle density, is convergent.
Setting $j_k=0$ in \eqref{eq:j_k} gives a recursion relation that determines $\tilde\rho_k$ from $\rho_1$,
\begin{equation}
  \tilde\rho_k = \rho_1^ke^{\frac{\eps_k}{k_BT}}, \text{ for } k\ge2.\label{eq:rho_k:equil}
\end{equation}
Here, we used $s_k=\rho_1$ for all $k$, since $\rho_1$ should be uniform in the equilibrium case. 
Substituting these $\rho_k$ into \eqref{eq:conservation} gives,
\begin{equation}
  \rho = \rho(\rho_1)\equiv \sum_{k=1}^{\infty}k\rho_1^ke^{\frac{\eps_k}{k_BT}}.\label{eq:conservation:equil}
\end{equation}
Hence, equilibria exist for monomer densities $\rho_1$ so that this series converges.
For large values of $k$, the binding energy can be written as 
\begin{equation*}
  \eps_k \approx k_BT\brk{\alpha k - \frac32\sig k^{2/3}}
\end{equation*}
Therefore, convergence happens for 
\begin{equation*}
  \rho_1 \le e^{-\alpha}=\rho_s.
\end{equation*}
In other words, equilibria exist only if the super-saturation is non-positive.
The largest value of  total particle density for which there is equilibrium is obtained by setting $\rho_1=\rho_s$ in \eqref{eq:conservation:equil}.
We denote this \emph{critical particle density} by $\rho_c$, 
\begin{equation}
  \rho_c=\rho(\rho_s).\label{eq:rho_c}
\end{equation}
Since $\rho_s\ll1$, the first few terms of the series give a close approximation of $\rho_c$.
\subsection{Zeldovich Nucleation Rate}
For positive super-saturation $\eta$, there is the critical cluster size $k=k_c$, for which the free energy cost to assemble a $k$-cluster from dissociated monomers is maximized. 
In the small super-saturation limit $\eta\goto0^+$, and $k\gg1$, it
follows from \eqref{eq:free:energy} and \eqref{eq:supersaturation},
that
\begin{equation}
  g_k \approx k_BT\brk{\frac32 \sig k^\frac23 -\alpha \eta},\label{eq:free:energy:2}
\end{equation}
and that the free energy cost (in units of $k_BT$) of the critical cluster is asymptotic to
\begin{equation*}
  g\equiv \max_{k} g_k \approx \frac{\sig^3}{{2\eta^2}}.
\end{equation*}
For small super-saturation, the free energy cost is high, and an analogy with the famous Arrhenius rate suggests that super-critical nuclei with $k>k_c$ are produced at a rate proportional to the exponential $e^{-g}$.
Since this proposed creation rate is exponentially small as $\eta\goto
0$, one might expect that after some initial transient, quasi-static
but non-equilibrium values of $\rho_k$ are established for $k$ on the
order of $k_c$, in which the discrete fluxes $j_k$ in
\eqref{eq:discrete:flux} are asymptotically equal to a uniform value $j$.
This $j$ is the creation rate of super-critical nuclei, proportional to $e^{-g}$. 
These essential ideas of nucleation kinetics are set forth in a famous paper of Zeldovich, on the nucleation of vapor bubbles for under pressurized liquid \cite{ZELD43}.
His starting point is a discrete system of kinetic ODE's like BD, but he first passes to a PDE limit of the ODE's and calculates the nucleation rate from the PDE\@.
Here, we implement the essential Zeldovich ideas, but within the framework of the discrete BD ODE's.

We work in the limit $k_c\ll k_*$, so for $k$ on the order of $k_c$ there is negligible difference between the surface value, $s_k$, of monomer density in \eqref{eq:s_k}, and the uniform value, $\rho_1$, far from clusters.
We show this: in \eqref{eq:s_k} for $s_k$, we see that $s_k\approx\rho_s(1+\eta_\infty) = \rho_1$ if $\oneover{\mu}\ll\eta_\infty$ and $\frac{k^{1/3}}{\sig\mu}\ll1$.
For $k_c\approx \brk{\sig/\eta_\infty}^3$ in \eqref{eq:critical:k:2} and $k_*=(\sig\mu)^3$ in \eqref{eq:k_*}, we find $\frac{k_c}{k_*}=\oneover{(\mu\eta_\infty)^3}$.
So, $\frac{k_c}{k_*}\ll 1$ implies $\oneover{\mu} \ll \eta_\infty$.
Furthermore, for $k$ on the order of $k_c$, $\frac{k^{1/3}}{\sig\mu}$ is on the order of $\brk{\frac{k_c}{k_*}}^{1/3}\ll1$.

In \eqref{eq:j_k}, we replace $s_k$ by $\rho_1$, and $j_k$ by $j$, to obtain a recursion equation that determines the $\rho_k$ for $k\ge2$ from $\rho_1$.
We write it as
\begin{equation}
  \frac{\rho_k}{\tilde\rho_k}-\frac{\rho_{k+1}}{\tilde\rho_{k+1}} = \frac{j}{d_k} e^{\frac{g_{k+1}}{k_BT}}.\label{eq:rho_k:quasi:stable}
\end{equation}
Here, $g_k$ is the free energy cost of a $k$-cluster, given in \eqref{eq:free:energy:2}, and $\tilde\rho_k$ denotes the solution \eqref{eq:rho_k:equil} of the homogeneous recursion relation with $j=0$, and $\tilde\rho_1=\rho_1$.
For positive super-saturation $\eta$, $\tilde\rho_k\goto\infty$ as $k\goto\infty$, and we expect that the $\rho_k$ in \eqref{eq:rho_k:quasi:stable} have $\frac{\rho_k}{\tilde\rho_k}\goto0$ as $k\goto\infty$.
Summing \eqref{eq:rho_k:quasi:stable} over $k$ gives a formula for $j$.
On the LHS, we get a telescoping sum with value 
\begin{equation*}
  \frac{\rho_1}{\tilde\rho_1}-\lim_{k\goto\infty}\frac{\rho_k}{\tilde\rho_k}=1-0=1.
\end{equation*}
Hence,
\begin{equation}
  1 = j\sum_{k=2}^{\infty}\oneover{d_{k-1}}e^{\frac{g_k}{k_BT}}.\label{eq:flux:sum}
\end{equation}
In the RHS, $g_k$ decreases linearly with $k$ as $k\goto\infty$, so the series converges.
In the small super-saturation limit $\eta\goto0^+$, we expect that the
sum on the RHS is dominated by terms  with $k$ near
$k_c\approx\brk{\frac{\sig}{\eta}}^3$, where $g_k$ attains its
maximum.
The relevant approximation to $g_k$ as $\eta\goto0^+$ and $k$ is on the order of $k_c$ is given by \eqref{eq:free:energy:2}.
Also, $d_k\approx \omega k^{2/3}$ as in \eqref{eq:d_k}.
Hence, \eqref{eq:flux:sum} has the asymptotic approximation 
\begin{equation}
  1 = \frac{j}{\omega}\sum_{k=2}^{\infty}k^{-\frac23}e^{\frac32\sig k^\frac23 - k\eta}.\label{eq:flux:sum:approx}
\end{equation}
The final step is the approximation of the sum by an integral, and evaluation of the $\eta\goto0^+$ limit by the saddle point method.
This leads to the approximation of $j$,
\begin{equation}
  j\sim\omega\sqrt{\frac{\sig}{6\pi}} e^{-\frac{\sig^3}{2\eta^2}}.\label{eq:zeldovich:rate}
\end{equation}
\section{Advection Signaling Problem}\label{sec:signaling:problem}
%
%
%


\begin{figure}[h!]
  \centerline{
\parbox{10cm} {\centerline{
\resizebox{8cm}{!}{\includegraphics{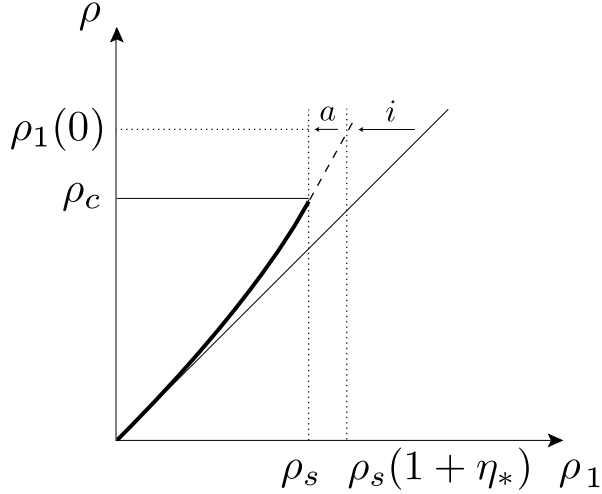}}}
\caption[The graph of $\rho(\rho_1)$ in
  $0\le\rho_1\le\rho_s$]{The darkened curve is the graph of $\rho(\rho_1)$ in
  $0\le\rho_1\le\rho_s$.
The dashed line is its linear interpolation into $\rho_a>rho_s$.
It intersects $\rho=\rho_1(0)>\rho_c$ at $\rho_1=\rho_s(1+\eta_*)$.
The arrow labeled $i$ represents the decrease of $\rho_1$ during
  ignition, and the arrow $a$ represents the decrease during the subsequent
  aggregation process.}\label{fig:aggregation:process} 
}}
\end{figure}

We examine the aggregation process, starting from a super-critical
density of particles, $\rho>\rho_c$, all in the form of monomers at
time $t=0$.
That is, 
\begin{equation*}
  \rho_1(0) = \rho>\rho_c,\quad \rho_k(0)=0 \text{ for } k\ge2.
\end{equation*}
There is an initial transient, called \emph{ignition}, in which the
first supercritical clusters appear. 
A detailed analysis of the ignition transient appears in a paper by
Bonilla {\it et al.} \cite{NBC05}.
Here we give a brief summary. First, sub-critical $(k<k_c)$ clusters are
created with quasi-static densities close to the values
$\tilde\rho_k$ in \eqref{eq:rho_k:equil}.
Of course, the value of $\rho_1$ becomes less than $\rho_1(0)=\rho$,
since these sub-critical clusters are created from monomers.
Hence, the appearance of the sub-critical quasi-static densities is
accomplished by the decrease of super-saturation from an initial value
of $\eta(0) = \frac{\rho-\rho_s}{\rho_s}$ to a smaller value, which we
denote by $\eta_*$. 
In appendix \ref{sub:sec:eta:eff}, we show that $\eta_*$ is related to $\rho-\rho_c$ by
\begin{equation}
  \rho-\rho_c \approx \eta_* \rho_s \rho'(\rho_s) \label{eq:eta:eff}
\end{equation}
as $\rho\goto\rho_c^+$.
Here, $\rho=\rho(\rho_1)$ is the equilibrium relation
\eqref{eq:conservation:equil} between $\rho_1$ and $\rho$ for
$0<\rho_1\le\rho_s$.
Figure~\ref{fig:aggregation:process} is a visualization of the
relation \eqref{eq:eta:eff} between $\eta_*$ and $\rho-\rho_c$.
The `completion' of the quasi-static densities for $1\le k< k_c=\brk{\frac{\sig}{\eta_*}}^3$ is
accompanied by the appearance of the first super-critical clusters
with $k>k_c$.
The rate of creation rises from zero to the Zeldovich rate in
\eqref{eq:zeldovich:rate} with $\eta=\eta_*$.
We assume $\eta_*$ is so that $k_c\ll
k_*$, so the Zeldovich rate indeed applies.

Now our focus shifts to the evolving distribution of the \emph{super-critical}
clusters.
The model we present here has three physical ingredients.
Two of them are the growth of the clusters, and their creation.
Both processes contain the super-saturation as a parameter, and the
remaining  ingredient is the connection of
super-saturation to the distribution of cluster sizes.

First, growth.
We expect a predominance of super-critical clusters with $k\gg k_*$ that undergo
diffusion limited growth.
Here is the heuristic argument for not resolving size scales
comparable to $k_*$ or smaller: once a cluster achieves super-critical
size, with $\frac{k-k_c}{k_c} \gg \eta^2$ (see the appendix \ref{sub:sec:det:BD})
it continues to grow  nearly deterministically, as shown in
section~3. 
Since the Zeldovich rate is exponentially small in $\eta$, an
exponentially long time elapses before the super-saturation shows an
significant decrease below the effective initial value, $\eta_*$,
established during ignition.
In this exponentially long time, we expect the super-critical clusters
to grow to sizes $k\gg k_*$, the regime of diffusion limited growth.

We assume that the characteristic cluster size $\bar k$ corresponding
to significant variations of the densities $\rho_k$ for $k\gg k_*\gg1$
is itself much larger than $k_*$, and this motivates a continuum
limit,
\begin{equation}
  \rho_k(t) \sim r(k,t).\label{eq:continuum}
\end{equation}
Here, $r(k,t)$ is a smooth function of its arguments, with the
characteristic scale $\bar k$ of $k$ much larger than $k_*$.
Substituting \eqref{eq:continuum} for $\rho_k$ into the discrete
advection-diffusion equations \eqref{eq:discrete:ad:dif} and using the
assumed largeness of $\bar k$, it follows that $r(k,t)$
asymptotically satisfies the advection PDE,
\begin{equation}
  \partial_t r + \partial_k (u\, r) =0. \label{eq:continuum:PDE}
\end{equation}
Here, the advection velocity $u=u(k,\eta)$ is identified from ODE
\eqref{eq:LS:dim} for  diffusion limited growth. 
We have,
\begin{equation}
 u(k,\eta)= d \brk{\eta \,k^\third -\sig},\label{eq:continuum:PDE:vel}
\end{equation}
where $\eta = \eta(t)$ is the `background' super-saturation, far from
any cluster.

Next, creation. 
In the analysis according to Zeldovich, recall that the discrete flux
$j_k$ in \eqref{eq:j_k} is asymptotically uniform for $k$ on the order
of $k_c$, with value $j$ given by \eqref{eq:zeldovich:rate}.
Here, we make the stronger assumption that the range of $k$ with the
asymptotically uniform value $j$ of $j_k$ extends to a scale of $k$
much larger than $k_*$ but smaller than the characteristic cluster
size $\bar k$ associated with the continuum limit
\eqref{eq:continuum}.
In this case, we expect an asymptotic matching between the continuum
limit of $j_k$, given by
\begin{equation}
  j_k \approx u(k,\eta) \, r(k,t),\label{eq:flux:BC}
\end{equation}
and the uniform value, $j$, of $j_k$, in some overlap domain of
cluster sizes $k$ much larger than $k_*$, but much smaller than $\bar
k$.
Since $k\gg k_*\gg k_c=\brk{\frac{\sig}{\eta_*}}^3$ in the overlap
domain, the dominant component of $u$ in \eqref{eq:continuum:PDE:vel}
is $d\,\eta\,k^{1/3}$ and \eqref{eq:flux:BC} reduces to
\begin{equation*}
  j_k \sim d\,\eta\,k^{1/3} r(k,t).
\end{equation*}
in the overlap domain.
Hence,  we propose the effective boundary condition on $r(k,t)$ on
$k=0$ are
\begin{equation}
   d\,\eta\,k^{1/3} \, r(k,t) \goto j = \omega\sqrt{\frac{\sig}{6\pi}}e^{-\frac{\sig^3}{2\eta^2}}\quad \text{ as } k\goto0.\label{eq:continuum:BC}
\end{equation}
If $\eta = \eta(t)$ is known, the advection PDE
\eqref{eq:continuum:PDE} and creation BC \eqref{eq:continuum:BC}
lead to a simple determination of $r(k,t)$, starting from the initial
condition of pure monomer, $r(k,0)\equiv 0$ in $k>0$. 

It remains to connect $\eta(t)$ to $r(k,t)$.
In the particle conservation identity \eqref{eq:conservation},
we now distinguish between sub-critical and super-critical clusters sizes.
That is,
\begin{equation*}
  \rho =\sum_{\makebox[0mm]{$\scriptstyle 1\le k\le k_c$}} k \rho_k + \sum_{k_c < k}k \rho_k.
\end{equation*}
The sub-critical sum is approximated by substituting equilibrium
values \eqref{eq:rho_k:equil} for $\rho_k$ based on super-saturation
$\eta$,
\begin{equation*}
  \rho_k\approx\tilde\rho_k = (1+\eta)^k\rho_s^ke^{\frac{\eps_k}{k_BT}},
\end{equation*}
and then taking the limit $\eta\goto 0$. 
The details here are a re-run of the calculation in appendix \ref{sub:sec:eta:eff}. 
We get
\begin{equation*}
  \sum_{k=1}^{k_c} k\,\rho_k \approx \rho_c + \eta \rho_s \rho'(\rho_s).
\end{equation*}
The super-critical sum is approximated by the integral,
\begin{equation*}
  \int_0^{\makebox[0mm][l]{$\scriptstyle\, \infty$}} k\,r(k,t)\,dk.
\end{equation*}
Here, the lower limit is $k=0$ and not $k_c$, because the
characteristic scale, $\bar k$, of $k$ in $r(k,t)$ is much larger than
$k_c$. 
In summary, the conservation identity \eqref{eq:conservation:equil} takes the
asymptotic form 
\begin{equation}
  \rho \approx \rho_c + \eta\rho_s\rho'(\rho_s)+\int_0^{\makebox[0mm][l]{$\scriptstyle\, \infty$}}
  k\,r(k,t)\,dk. \label{eq:continuum:conserv}
\end{equation}
Notice that if $r\equiv0$, which corresponds to negligible
super-critical clusters, \eqref{eq:continuum:conserv} reduces to \eqref{eq:eta:eff} with
$\eta=\eta_*$.
Hence, $\eta(t)$ has the effective initial condition $\eta(0) =\eta_*$.

Thus, the signaling problem for $r(k,t)$ consists of the advection PDE
\eqref{eq:continuum:PDE}, the creation BC \eqref{eq:continuum:BC}, and the functional dependence of
the super-saturation, $\eta$ on $r(k,t)$ in \eqref{eq:continuum:conserv}.
This signaling problem is nonlinear because $\eta$, as a functional of
$r$, appears in the advection PDE and, in addition, in the exponential
creation rate in the BC\@. 
\section{Conclusions}
The discrepancy between the two accepted model of nucleation---the surface reaction
model derived by Becker and D\"oring, and the diffusion limited growth
model due to Lifshitz and Slyozov---has now been resolved.
Our new model, in which  clusters interact with diffusing monomers 
by a surface reaction, predicts both models as limit cases. 
Although it provides a growth rate for clusters of all sizes, 
from small to large,  its limiting behaviors are of special interest for us. 
In the limit of small clusters, the BD model emerges, and we can derive
the Zeldovich creation rate of super-critical clusters using the BD kinetics.
In the limit of large clusters, the diffusion limited growth model
emerges, from which we derive the same evolution PDE for the density
function, $r(k,t)$, as LS.

Not all our findings corroborate the classical assumptions and results.
We find that the `common wisdom' about the  monomer
density at the cluster surface is wrong. 
In the classical DLG model, the surface density is chosen so that the
surface reaction is in equilibrium. 
In our model, the balance between the surface reaction and 
diffusion leads to a surface monomer density substantially different
from the `equilibrium' value.  
Despite this, the new model still predicts the same growth rate as the
classical DLG model (for large clusters).

Another piece of `common wisdom' is that BD pertains to small clusters and DLG to large ones. 
In our model, there is a new characteristic cluster size, $k_*$, that
separates small ($k\ll k_*$) from large ($k\gg k_*$). 
This is useful in providing definite predictions for the validity of
the model. 
For example, since nucleation happens around the critical size $k_c$, and the
Zeldovich creation rate assumes the small cluster limit of the model, the
validity of the Zeldovich formula requires $k_c\ll k_*$. 
The case where $k_c$ is of the same order or larger than $k_*$ is not
covered by the current thesis nor by classical Zeldovich nucleation
theory. 
A separate investigation is required for this case, one that goes beyond  or treatment of the  monomer bath as a smooth `mean field' and really accounts for its discrete and fluctuation nature.

The monomer density around the cluster satisfies  the diffusion
equation and a mixed boundary condition at the surface of the
cluster. 
To calculate the flux of the monomers into or out of the cluster,
our model assumes that the monomer density in the vicinity of the
cluster is a quasi-static solution of the diffusion equation.
We show that the quasi-static `surface' value of monomer concentration
in our model is stable: starting from general initial conditions, the concentration of the monomer bath rapidly relaxes to the quasi-static approximation.

The preceding insights are codified in a \emph{signaling problem} for the cluster size distribution.
It consists of an advection PDE consistent with diffusion limited growth, and a BC consistent with the Zeldovich creation rate.
The PDE and BC contain the super-saturation as a parameter.
The super-saturation, in turn, is a function of the cluster size distribution as dictated be the overall conservation of particles.
Hence, the full signaling problem is non-linear. 
In particular, the Zeldovich creation rate is \emph{exponentially} small as the super-saturation goes to zero.
Therefore, accounting for small corrections to the monomer density is
important, especially in the initial stage when new clusters are
being created. 

Given an initial condition of pure monomers, quasi-static densities  are
formed during the ignition phase\cite{NBC05}, whereby small,
sub-critical clusters are created from the monomers. 
The ignition transient is a precursor to the nucleation of
super-critical clusters.
Therefore, once the nucleation `has ignited', the  monomers density is 
lower than the its original value, prior to the ignition transient.
The signaling problem starts from an effective IC that explicitly accounts for the loss of monomers during ignition.
This is especially relevant  in our case, where the
monomer-density is close to the saturation density, $\rho_s$, and the dependence of the PDE and BC on the monomer density is very strong. 

Such is the state of the \emph{theory}.
We now conclude the conclusion by a small trespass into the domain of
accountability: do the material parameters of real aggregation
processes cooperate with the various assumptions of the modeling?
We start with the characteristic cluster size $k_*$, most
conspicuously present in our model.
We examine physical constants associated with an aqueous solution of calcium
carbonate, CaCO\subs{3}.
Its solubility is rather small, so the basic requirement of
`diluteness' is satisfied.
Admittedly, CaCO\subs{3} dissolves into positive (Ca\sups{+}) and
negative ($\text{CO}^-_3$) ions, which is not reflected  in our
aggregation model with identical particles.
But we are examining \emph{crude} order of magnitude estimates, so
will not be distracted by the inconvenience, and we formally consider
a positive-negative ion pair as a `monomer.'

Formula \eqref{eq:k_*} for $k_*$ contains molecular volume $\nu$, the
saturation density of monomer $\rho_s$, the diffusion coefficient $D$ of
monomer in solution, and the dissociation rate constant $\omega$.
Estimates of $\nu, \rho_c,$ and $D$ are readily found in a
chemical handbook \cite{CRCTable}.
We use $\rho_c$ as an order of magnitude estimate of $\rho_s$.
The dissociation rate, $\omega$, is much more elusive.
Here, we indulge in the activation energy model similar to Kelton's,
in his review of glass-crystal transitions \cite{KELT91}.
The model is summarized by the formula
\begin{equation}
  \omega = \frac{D}{\nu^{2/3}}e^{-\beta},\label{eq:kelton:formula}
\end{equation}
Here, $\frac{\nu^{2/3}}{D}$is the characteristic time for a monomer to
diffuse a distance comparable to its own size, and $\beta$ is an
`activation energy of dissociation,' in units of $k_BT$.
Inserting \eqref{eq:kelton:formula} for $\omega$ into \eqref{eq:k_*},
we find
\begin{align}
  k_*&=(3\cdot 16\pi^2) \brk{e^\beta\rho_s}^3,\notag
\intertext{or, using $\rho_s=e^{-\alpha}$,}
 k_*&=(3\cdot 16\pi^2) e^{3\brk{\beta-\alpha}}.\label{eq:k_*:2}
\end{align}

Formula \eqref{eq:kelton:formula} has other  applications for us,
besides the estimate \eqref{eq:k_*:2} of $k_*$.
For instance, recall that the quasi-static limit of the FBP \eqref{eq:eta:lap} is based on the `diffusion time' $\frac{\nu^{2/3}}{D}$ much smaller than the dissociation time $\frac{1}{\omega}$, so we required $\frac{D}{\omega\nu^{2/3}}\gg1$.
From \eqref{eq:kelton:formula} we get 
\begin{equation}
  \frac{D}{\omega\nu^{2/3}} = e^\beta\label{eq:beta:ratio}
\end{equation}
and so an activation energy $k_BT\beta$ on the order of a few $K_BT$ is sufficient.

Next, we examine the criterion $\frac{k_c}{k_*}\ll1$ for the validity of the Zeldovich nucleation rate.
Using (\ref{eq:critical:k:2}, \ref{eq:k_*}), we find this implies a bound on the super-saturation,
\begin{equation*}
  \eta\gg \sig \,e^{\alpha-\beta}.
\end{equation*}
Using the crude `cube' model of bonding energy \eqref{eq:energy:cube} we estimate $\sig$ by, $\sig\approx \frac23\alpha$ so the criterion on 
the supersaturation becomes
\begin{equation}
  \eta \gg \alpha \,e^{\alpha-\beta}.\label{eq:eta:lower:bound}
\end{equation}

A lower bound on the super-saturation might seem like a problem, as we expect $\eta$ to asymptote to zero in the long-time limit of an aggregation process.
However, significant nucleation occurs at an early and relatively brief phase, so \eqref{eq:eta:lower:bound} should apply for the initial super-saturation. 
In late stage coarsening, $\eta$ is  much smaller than the RHS of \eqref{eq:eta:lower:bound}, but significant nucleation is not happening then.

Numerical evaluation of $k_*$, or the RHS of \eqref{eq:eta:lower:bound} require actual values of $\alpha$ and $\beta$.
It is relatively easy to find $\alpha$, as we have $\alpha = -\log \rho_s$ and $\rho_s$ is essentially the volume fraction of monomer in saturated solution.
For instance, from the data in \cite{CRCTable}, we find that $4.1\times10^{-4}\text{cm}^3$ of solid CaCO\subs{3} is soluble in 100cm\sups{3} of water at room temperature. 
The volume of the solution is nearly 100cm\sups{3} so the volume fraction occupied by Ca\sups{+}, $\text{CO}^-_3$ is (roughly) $\rho_s=4.1\times10^{-6}$, and our estimate of $\alpha$ is $\alpha\approx7.8$.
The bad news is that we do not know $\beta$ any better than we know $\omega$ in the first place, so we cannot do primia-facie evaluations of $k_*$ or the RHS of \eqref{eq:eta:lower:bound}.

Our policy is to use  \eqref{eq:eta:lower:bound} to obtain bounds on $\beta$ for which our model is valid.
For instance, we have seen that the quasi-static approximation requires $\beta\gg1$.  
In addition, our whole analysis is based on small super-saturation, so $\eta\ll1$.
This is compatible with \eqref{eq:eta:lower:bound} only if 
\begin{equation}
  \frac{\alpha}{e^{\beta-\alpha}}\ll1.\label{eq:beta:bound:2}
\end{equation}
Starting with $\alpha=7.8$, we find the LHS is unity if
$\beta-\alpha=2$, but if we increase the activation energy, $\beta$, by $3\,k_BT$ we get $\beta-\alpha=5$ and  the LHS of \eqref{eq:beta:bound:2} is $0.05\ll1$.
Inserting this value of $\beta$ into \eqref{eq:k_*:2} for $k_*$ results in $k_*\approx1.5\times10^9$.

\chapter{The Three Eras}

%
%
%
\section{Introduction}
Aggregation of identical particles (monomers) into
clusters is a universal phenomenon throughout physics, chemistry and
biology  \cite{KGT83}, \cite{KELT91}, \cite{NB00}, \cite{GWS01},
\cite{ISRAELACHVILI91}, \cite{NCB02}, \cite{LS61}, \cite{XH91},
\cite{MG96}, \cite{GNON03}. 
There are two classical models. 
The Becker-D\"oring (BD) theory \cite{BD35}, which
is based on the `surface reactions' of adding or subtracting one particle
(monomer) at a time from the surface of a cluster. 
In the Lifshitz-Slyozov (LS) theory of diffusion-limited growth (DLG)
\cite{LS61} a gradient of monomer concentration surrounding a cluster
supports a diffusive influx of monomers into it. 
It is generally accepted that BD models the early stage of nucleation,
in which `super-critical' clusters are created according to the
Zeldovich nucleation rate, and that DLG models the subsequent growth
of the large clusters. 
In the previous chapter, we present a more general
model that contains BD and LS as limits of small and large cluster
sizes, respectively.
A continuum limit of this expanded model yields a \emph{signaling
  problem} for evolving the distribution of clusters in the space of
their sizes.
The cluster distribution satisfies an advection PDE, with the
advection velocity specified according to DLG, and an effective
boundary condition (BC) at zero which represents the creation of
clusters at the Zeldovich rate.
Both the advection PDE and `creation' BC contain the excess monomer
concentration, the \emph{super-saturation}, seen far from clusters as
a parameter, and the super-saturation is, in turn, a functional of the
cluster size distribution as dictated by the overall conservation of particles.
This is the source of strong nonlinearities in the signaling problem,
nonlinearities that lead to the novel separation of scales behind the
`three era' mentioned in the abstract.
The analysis of each era consists of finding convenient variables and
scales, deriving the reduced equations and finally solving the
equations. 
As a result, we have the relevant \emph{physical} scales of each of
the eras. 

The chapter is structured as follows.
In section 2 we formulate the signaling problem.
The governing equations and the perturbation parameter (the initial super-saturation) are introduced. 
In sections 3--5 we resolve the three distinguished limits that
correspond to the eras: `nucleation', 'growth' and 'coarsening'. 
Effective initial conditions for the growth era follow from matching
with the `tail' of the nucleation era, and the tail of the growth era
similarly provides initial conditions for the coarsening era.

Section 3 is the analysis of the nucleation era.
From the reduced equations we derive a single integral equation for
the super-saturation as a function of time.
From the numerical solution of the integral equation we readily obtain
the evolving cluster size distribution.
In section 4, we characterize the cluster distribution during the
growth era by ODE's for its mean and deviation. 
The distribution is `narrow' because the mean size is large
compared to the deviation, so at any given time during the growth era
we have an ensemble of clusters all with nearly the same size.
The growth slows down when the super-saturation is depleted to a
sufficiently low level, and the mean cluster size seems to stabilize. 
However, the deviation continues to grow slowly (relative to the
characteristic time of the growth era).
This is the precursor to the coarsening era.
Initially, we follow the slow widening analytically, by using a
linearized approximation to the advection velocity over the (still
narrow) support of the cluster size distribution.
Later as this approximation becomes less valid due to the widening of
the support, the solution to the reduced equations of the coarsening
era is followed numerically.
The numerical solution displays full-blown coarsening. 
Smaller clusters dissolve back into monomers, which are then adsorbed
by the surviving large clusters.
The numerical coarsening era solution eventually relaxes to a unique
self-similar distribution, selected from the one parameter family of
similarity solutions to the original equations of LS.

\section{The Signaling Problem}
The detailed physical-mathematical reasoning behind our reduced model
of nucleation and coarsening is the subject of the previous chapter. 
A summary of essentials, given in the current section, serves as the starting point of the current chapter.

First, we identify the initial conditions that lead to aggregation, as
opposed to equilibrium with negligible densities of large clusters.
\emph{Equilibrium}, if achievable, is characterized by densities
$\rho_k$ of $k$-clusters given by
\begin{equation}
  \rho_k=\tilde\rho_k\equiv \rho_1^ke^{\frac{\eps_k}{kBT}},\quad k\ge 2.\label{eq:rho_k:equil:kin}
\end{equation}
Here, all densities are in units of $\oneover{\nu}$, where $\nu$ is
the molecular volume and $\eps_k$ is the binding energy of a
$k$-cluster: the energy cost to dissolve it into separated monomers.
Of course, $\eps_1=0$. 
For $k\gg1$ we use the standard model,
\begin{equation}
\eps_k \sim k_BT\brk{\alpha k-\frac32\sig k^{\frac23} }.\label{eq:energy}
\end{equation}
The positive constants $\alpha$ and $\sig$ are prefactors of terms
proportional to the volume and surface area of the cluster.
The total density of particles is
\begin{equation}
  \rho=\sum_{k=1}^\infty k \rho_k.\label{eq:total:density}
\end{equation}
Substituting the equilibrium densities \eqref{eq:rho_k:equil:kin} into
\eqref{eq:total:density} gives $\rho$ as as power series in monomer
density $\rho_1$.
It converges for $0\le\rho_1\le \rho_s$, where $\rho_s$ is the
\emph{critical monomer density}
\begin{equation}
  \rho_s\equiv e^{-\alpha}.\label{eq:critical:mono}
\end{equation}
Hence,
\begin{equation}
  \rho = \rho(\rho_1) \equiv \sum_{k=1}^\infty e^{\frac{\eps_k}{k_BT}}\rho_1^k\label{eq:rho}
\end{equation}
in $0\le\rho_1\le\rho_s$.
The total particle density, $\rho(\rho_1)$ is an increasing function.
Its maximum value, achieved at $\rho_1=\rho_s$, is the \emph{critical
  particle density}, $\rho_c\equiv\rho(\rho_s)$.

Since the first term of the sum in \eqref{eq:total:density} is
$\rho_1$ and the rest of the terms are positive, it follows that
$\rho_c>\rho_s$.
If we start from pure monomer with $0<\rho_1\le\rho_c$, we expect that
the densities $\rho_k=\rho_k(t)$ relax to equilibrium values
\eqref{eq:rho_k:equil:kin} as $t\goto\infty$.
If initially $\rho_1>\rho_c$, we expect aggregation and coarsening.

During an aggregation process, the \emph{super-saturation} 
\begin{equation*}
  \eta \equiv \frac{\rho_1-\rho_s}{\rho_s}
\end{equation*}
decreases from the positive initial value $\frac{\rho-\rho_s}{\rho_s}$
towards zero as $t\goto\infty$.
Positive super-saturation is a `driving' parameter of aggregation.
If $\eta<0$, all clusters have strong tendencies to dissolve into
monomer, independently of their size.
If $\eta>0$ there is a \emph{critical size}
\begin{equation*}
  k_c\sim\brk{\frac{\sig}{\eta}}^3
\end{equation*}
so that \emph{sub-critical} clusters with $k<k_c$ still have a strong
tendency to shrink, but \emph{super-critical} clusters with $k>k_c$
have a strong tendency to persist and grow.
The essential hurdle to get aggregation going is a long and highly
unlikely sequence of `favorable fluctuations' whereby an initially
sub-critical cluster eventually exceeds critical size and grows.
An analysis along the lines of Zeldovich \cite{ZELD43} shows that the
rate of creation of super-critical clusters per molecular volume $\nu$
is asymptotic to 
\begin{equation}
  j\sim \Omega e^{-\frac{\sig^3}{2\eta^2}},\quad \Omega\equiv \omega\sqrt{\frac{\sig}{6\pi}},\label{eq:zeldovich}
\end{equation}
as $\eta\goto0^+$.
Here, $\omega$ is the rate constant for a single particle to dissolve
into the monomer bath from the surface of a cluster.
We see that the creation rate is exponentially small as
$\eta\goto0^+$.
Nevertheless, it is not zero.
Aggregation \emph{is} going to happen, but it might take a long time.

There are three essential ideas in our reduced model of aggregation:
transport of clusters in the space of their size, their creation at
the Zeldovich rate \eqref{eq:zeldovich}, and the functional dependence
of super-saturation on the distribution of cluster sizes as dictated
by overall conservation of particles.
The latter is essential, because the super-saturation has a strong
effect upon both the creation rate and the transport of clusters.

Transport is modeled by \emph{diffusion limited growth} (DLG) in which
the size of any particular cluster is treated as a deterministic
function of time, $k=k(t)$, which satisfies the ODE
\begin{equation}
  \dot k = u(k,\eta)\equiv d(\eta\,k^{\third}-\sig).\label{eq:vel}
\end{equation}
The prefactor $d$ is proportional to the diffusion coefficient of
monomer,
\begin{equation*}
   d= \brk{3\cdot 16 \pi^2}^\frac13\frac{\rho_sD}{\nu^\frac23}. 
\end{equation*}
To describe an ensemble of clusters undergoing DLG, we introduce a
continuum approximation of the cluster densities.
\begin{equation}
  \rho_k(t)\sim r(k,t)
\end{equation}
for $k\gg1$. 
Here, $r(k,t)$ is a smooth function of its arguments.
We propose that $r(k,t)$ satisfies the advection PDE consistent with
individual cluster sizes evolving according to the ODE 
\eqref{eq:vel}.
That is,
\begin{equation}
  \partial_tr+\partial_k(u\,r)=0. \label{eq:pde}
\end{equation}
In the previous chapter, we show that DLG emerges from a
full model, which contains both surface reaction and monomer diffusion, in the
large cluster limit 
\begin{equation}
  k\gg k_*\equiv(3\cdot16\pi^2)^\third \frac{D\rho_s}{\omega \nu^{\frac23}}.\label{eq:k_*:kin}
\end{equation}
In all this, we are tacitly assuming a preponderance of $k\gg k_*$
clusters, all undergoing DLG\@.
The argument for this is related to the creation of super-critical
clusters, which we take up now.

The creation of super-critical clusters at the Zeldovich rate \cite{ZELD43} is represented by an effective BC on $r$ at $k=0$:
\begin{equation}
  d\,\eta \,k^\third \,r \goto \Omega e^{-\frac{\sig^3}{2\eta^2}},  \text{ as } k\goto 0. \label{eq:zeldovich:BC}
\end{equation}
The heuristic argument behind \eqref{eq:zeldovich:BC} is given in the
previous chapter.
Here, we make two notes: 
First, the BC \eqref{eq:zeldovich:BC} is shorthand for an asymptotic equality of the Zeldovich rate \eqref{eq:zeldovich} and the advection flux $u\,r$ in an `intermediate' range of $k$, much larger that the critical size $k_c$, but much smaller than the characteristic size of $k$ associated with the distribution $r(k,t)$.
Notice that $d\,\eta\,k^\third$, which appears in the LHS of \eqref{eq:zeldovich}, is the dominant component of $u$ in $k\gg k_c$.
Second, an exponentially small creation rate as $\eta\goto0^+$ leads to an exponentially long time interval of significant nucleation.
The time for a single super-critical cluster to grow to size $k\gg k_*$ is presumably much shorter.
This is why we expect a preponderance of clusters in the DLG range of size, $k\gg k_*$.

The final ingredient of the reduced aggregation model is conservation of particles, which implies that the super-saturation is a functional of the cluster size distribution $r(k,t)$. 
After an initial transient, in which quasi-static  densities of sub-critical clusters ($k<k_c$) are established, the conservation is expressed by 
\begin{equation}
  \rho\sim \rho_c + \rho_s\rho'(\rho_s)\eta + \int_0^\infty k\,r(k,t)\,dk. \label{eq:conserv}
\end{equation}
Here, $\rho(\rho_1)$ is the equilibrium relation \eqref{eq:rho}.
In the RHS, $\rho_c+\rho_s\rho'(\rho_1)\eta$ approximates the total density of particles in sub-critical clusters in the $\eta\goto0^+$ limit.
The integral approximates the density of particles in super-critical clusters.
We rewrite \eqref{eq:conserv} as
\begin{equation*}
  \eta\sim \eta_* -\Lambda \int_0^\infty k\,r(k,t)\,dk,\quad \Lambda\equiv \oneover{\rho_s\rho'(\rho_s)}. \label{eq:conserv2}
\end{equation*}
Here $\eta_*$ is an \emph{effective} initial value of the super-saturation, 
\begin{equation*}
  \eta_*\equiv\frac{\rho-\rho_c}{\rho_s\rho'(\rho_s)}.
\end{equation*}
This value is achieved after the quasi-static, sub-critical densities are established, but before there is a significant number of particles in super-critical clusters.

The asymptotic analysis of the solution uses $\eta_*$ as a gauge parameter. 
That is, we define 
\begin{equation*}
\eps\equiv\eta_*
\end{equation*}
and scale $\eta$ with $\eps$. 
Replacing $\eta$ with $\eps\eta$ yields the signaling problem (SP)
\begin{align}
  \partial_tr+\partial_k(vr)&=0,\quad \text{ in } k>0,\label{eq:pde1}\\
  v &=d(k^\third\eps\eta-\sig),\label{eq:vel1}\\
  d\, \eps\, \eta k^\third \,r&\goto\Omega  e^{-\frac{\sig^3}{2\eps^2\eta^2}}, \quad\text{ as }
  k\goto0^+,\label{eq:BC:zeld}\\
 \eta &= 1 - \eps^{-1}\Lambda \int_0^\infty
  k\,r(k,t)\,dk. \label{eq:conserv1}
\intertext{In the following sections we determine the $\eps\goto0$
  asymptotic solution of the SP (\ref{eq:pde1}--\ref{eq:conserv1})
  with IC}
  r(k,0)&\equiv 0.\label{eq:IC1}
  \end{align}

\section{The Nucleation Era}
During the nucleation era, almost  all of the supercritical clusters are created.
It lasts until the Zeldovich creation rate decreases to a small fraction of its initial value.
First, we find the relevant scales of the variables from 
 dominant balances in (\ref{eq:pde1}--\ref{eq:conserv1}).
The reduction of  PDE \eqref{eq:pde1} has simple
characteristics that can be used to translate the SP into an integral
equation for $\delta\eta(t)$, the change of the super-saturation from
its initial value.
Finally, using the numerical solution for $\delta\eta(t)$ and the characteristics, we reconstruct the density of cluster sizes, $r(k,t)$.
\subsection{Nucleation Scalings}
For small $\eps$ the nucleation rate
\eqref{eq:BC:zeld} is highly sensitive to small relative changes in the super-saturation. 
We therefore expect that a small decrease in the super-saturation
causes it to decrease sharply. 
Instead of working directly with the super-saturation, $\eta$,
we work with the \emph{change} in super-saturation,
\begin{equation}
  \delta\eta=\eta-1.
\end{equation}
We find scaling units $[t],\, [k],\, [r],\, [\delta \eta]$ of the variables $t,\,k,\,r,\,\delta\eta$ from the dominant balances in
equations (\ref{eq:pde1}--\ref{eq:conserv1}).  
Since the nucleation rate is exponentially small in $\eps$, we expect that a long time
elapses before the super-saturation decreases enough to shut down the
creation of additional clusters.
During this long time the clusters grow to exponentially large sizes. 
Thus, the dominant term in the advection velocity in \eqref{eq:vel1}
is $k^\third\eps\eta$ and not $\sig$. 
The dominant balance associated with \eqref{eq:vel1} is therefore
\begin{equation}
  \Brk{k}\Brk{t}^{-1}=d\eps\Brk{k}^\third\label{eq:scale1}
\end{equation}
Integrating \eqref{eq:pde1} with respect to $k$, from $0$ to $\infty$ and use of BC \eqref{eq:BC:zeld} implies 
\begin{equation}
  \frac{d}{dt}\int_0^\infty  r(k,t)dk=\Omega e^{-\frac{\sig^3}{2\eps^2\eta^2}}\label{eq:PDE:intgrated}.
\end{equation}
Recalling that we expect $\eta$ to remain near 1, we see that the dominant  balance associated with \eqref{eq:PDE:intgrated} is
\begin{equation}
\Brk{t}^{-1}\Brk{r}\Brk{k} = \Omega e^{-\frac{\sig^3}{2\eps^2}}\label{eq:scale2}
\end{equation}
The conservation equation \eqref{eq:conserv1} can be written in terms of
$\delta\eta$,
\begin{equation*}
  \eps\delta\eta =-\Lambda \int_0^\infty k\,r(k,t)\,dk.
\end{equation*}
Therefore, the dominant balance associated with the conservation equation \eqref{eq:conserv1} is
\begin{equation}
 \eps[\delta\eta]= \Lambda\Brk{k}^2\Brk{r}.\label{eq:scale3}
\end{equation}
This gives us three dominant balance equations relating the four scales of the
problem.
To find unique scalings, one more equation is needed. 
This last equation identifies the change in super-saturation
$\delta\eta$ that gives rise to a  significant relative decrease in the Zeldovich nucleation rate \eqref{eq:BC:zeld}.
For $\delta\eta\ll\eta$, the change in the exponent
$-\frac{\sig^3}{2\eps^2\eta^2}$ in \eqref{eq:BC:zeld} is approximately
$\frac{\sig^3}{\eps^2\eta^3}\delta\eta$, so the relative change in the
nucleation rate, $\frac{\delta j}{j}$ is 
\begin{equation*}
  \frac{\delta j}{j} \sim e^{\frac{\sig^3}{\eps^2\eta^3}\delta\eta}.
\end{equation*}
Since $\eta$ is already rescaled to be near 1, in order to have an $\O(1)$ relative change in $j$ we must have
\begin{equation}
  \Brk{\delta\eta} = \frac{\eps^2}{\sig^3}.\label{eq:scale4}
\end{equation}
The solutions of  equations
(\ref{eq:scale1}, \ref{eq:scale2}, \ref{eq:scale3}, \ref{eq:scale4}) for the scaling units $\Brk{r}, \Brk{k}, \Brk{t},\Brk{\delta\eta}$ are summarized in the scaling table:
\begin{center}
Scaling Table\\
\begin{tabular}{l*{5}{>{$}c<{$}}}
Variable&k&t&r&\delta\eta\\
\hline\\[-4mm]
Unit
&\brk{\frac{d\eps^4}{\Omega\sig^3\Lambda}
  e^{\frac{\sig^3}{2\eps^2}}}^{\frac35}
&\brk{\frac{\eps}{d\sig^2}
  }^{\frac35} \brk{\Omega\Lambda}^{-\frac25}e^{\frac{\sig^3}{5\eps^2}}
&\brk{\frac{\Omega^2\sig\Lambda^{1/3}}{d^2\eps^3}
  e^{-\frac{\sig^3}{\eps^2}}}^\frac35
&\frac{\eps^2}{\sig^3}
\end{tabular}
\end{center}
The scaled versions of  (\ref{eq:pde1}--\ref{eq:conserv1}) are
\begin{align}
   \partial_tr+\partial_k( u r)=&0,\quad \text{ in } k>0,\label{eq:K:scaled:exact}\\
   u=& k^{\third} (1+\frac{\eps^2}{\sig^3}\delta\eta)-s,\label{eq:tag:s}\\
   \brk{1+\frac{\eps^3}{\sig^3}\delta\eta} k^{\third}r\goto & \, \exp\brk{-\frac{\sig^3}{2\eps^2}\brk{\frac{1}{\brk{1+\frac{\eps^2}{\sig^3}\delta\eta}^2}-1}},\quad \text{ as } k\goto0^+,\label{eq:zeld2}\\
    \delta\eta =& -\int_0^\infty kr\,dk,\label{eq:conservation:kin}
\end{align}
In \eqref{eq:tag:s}, $s$  is the exponentially small
combination of parameters
\begin{equation}
  s\equiv \brk{\frac{d^6\eps^9}{\Omega\sig^8\Lambda}
  e^{\frac{\sig^3}{2\eps^2}}}^{-\frac15}.
\end{equation}
In \eqref{eq:zeld2}, the exponent reduces to $\delta\eta$ as $\eps\goto\infty$, and this `confirms' the unit $\frac{\eps^2}{\sig^3}$ of $\delta\eta$ in the scaling table.
\subsection{Reduced Nucleation Kinetics}
%
%
\begin{figure}[ht!]
  \centerline{
\parbox{8cm}{\centerline{
\resizebox{6cm}{!}{\includegraphics{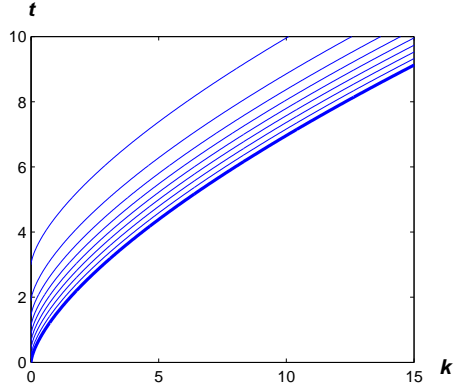}}}
\caption[Characteristic during the nucleation era]{The characteristics of the scaled PDE\@. The flux $k^{\third}r$ is
 constant along the characteristics. The density of the curves matches
 the value of the solution $r(k,t)$.}\label{fig:char}
}}
\end{figure}
In the limit $\eps\goto0$, equations (\ref{eq:K:scaled:exact}--\ref{eq:conservation:kin}) and the initial condition \eqref{eq:IC1} reduce to
\begin{align}
   \partial_tr+\partial_k(k^\third r)=&0,\quad \text{ in } k>0,
\label{eq:scaled:pde}\\
    k^{\third}r\goto & e^{\delta \eta},\quad \text{ as } k\goto0^+,\label{eq:BC1}\\
    \delta\eta =& -\int_0^\infty kr\,dk, \label{eq:nucleation:conservation}\\
    r(k,0) \equiv & 0.\label{eq:bc:r}
\end{align}
We solve the reduced SP (\ref{eq:scaled:pde}--\ref{eq:bc:r}) by
transforming it  into an integral equation for the change in super-saturation, $\delta\eta(t)$.
This integral equation is solved numerically, and the solution for $r(k,t)$ is recovered from $\delta\eta(t)$.

The flux of super-critical clusters, $j \equiv k^\third r$,
is constant along the characteristics 
\begin{equation}
  k = \brk{\frac23(t-\phi)}^\frac32, \text{ for } t\ge\phi \label{eq:world:line}
\end{equation}
of the PDE \eqref{eq:scaled:pde}.
The characteristics can be seen in Figure~\ref{fig:char}. 
In the figure, the (horizontal) density of the characteristics
indicates  the value of the distribution of cluster sizes at time
$t$.
Physically, each curve in \eqref{eq:world:line} describes the
world-line of a cluster nucleated at time $t=\phi$.
The region below the thick line in Figure~\ref{fig:char} corresponds to $t<\frac32k^\frac32$; places where the first nucleated clusters have yet to arrive.
Hence, the value of $r(k,t)$ there is zero.

For a \emph{given} $\delta\eta(t)$, the solution, $r(k,t)$, which follows from $k^\third r$ being constant along characteristics and the BC \eqref{eq:BC1}, is
\begin{equation}
  r(k,t)= \left\{
\begin{aligned}
&k^{-\third}e^{\delta\eta(t-\frac32k^\frac23)}, &t \ge \frac32k^\frac23,\\
&0, & 0\le t <\frac32k^\frac23.
\end{aligned}
\right.
 \label{eq:char:sol}
\end{equation}
The integral equation for $\delta\eta(t)$ emerges by substituting \eqref{eq:char:sol} for $r(k,t)$ into the conservation identity \eqref{eq:nucleation:conservation}. 
We get
\begin{align}
   \delta\eta(t) &= -\int_0^t\brk{\frac23(t-\phi)}^\frac32e^{\delta\eta(\phi)}\,d\phi.\label{eq:integral_1}
\end{align}
Notice that the variable of  integration in \eqref{eq:integral_1}  has been changed from $k$ in \eqref{eq:nucleation:conservation} to \mbox{$\phi\equiv t-\frac32k^{2/3}$} as determined by \eqref{eq:world:line}.
\subsection{Physical Predictions}
%
%


\begin{figure}[ht!]
  \centerline{
\parbox{8cm}{\centerline{
\resizebox{6cm}{!}{\includegraphics{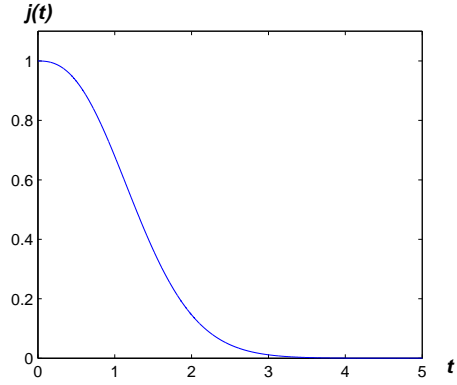}}}
\caption[Nucleation rate as a function of time]{The Zeldovich flux rate $j$ as a function of time. After $t=\O(1)$,
  the super-saturation
  decreases slightly, and the flux gets turned off.}
\label{fig:flux}
}}
\end{figure}

\begin{figure}[ht!]
  \centerline{
\parbox{8cm}{\centerline{
\resizebox{6cm}{!}{\includegraphics{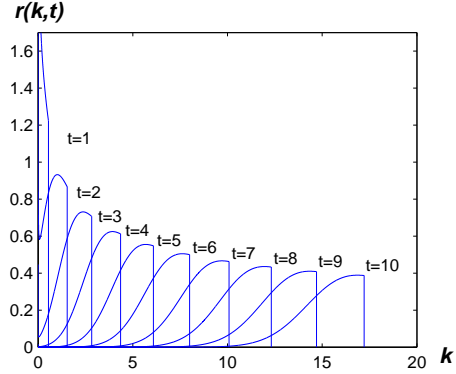}}}
\caption{The density of cluster sizes, $r(k,t)$, for
  various values of $t$.}
\label{fig:movie}
}}
\end{figure}

We solve (\ref{eq:integral_1}) numerically.
Figure~\ref{fig:flux} shows the flux as a function of time.
At $t$ near 1 the flux is half its original value,
and that at $t=5$ it has effectively vanished. 
A short discussion of the method and numerical result can be found
in Appendix \ref{sec:numerical:explain}.
Once $\delta\eta(t)$ is determined, we recover the distribution $r(k,t)$ of cluster sizes from \eqref{eq:char:sol}.
Figure~\ref{fig:movie} displays $r$ vs. $k$ for a sequence of times $t$.

We present collateral predictions for the total density of  clusters
generated during the nucleation era, the characteristic cluster size and the characteristic time for their nucleation. 
The (scaled) density of super-critical clusters is given by
\begin{equation*}
  \sum_{k>k_c}\rho_k \sim \int_0^\infty r\,dk.  
\end{equation*}
Using the change of variables from $k$ to $\phi$ according to
\eqref{eq:world:line}, and the PDE \eqref{eq:scaled:pde}, we convert 
 the integral of $r$ into an integral of the flux $j$, 
\begin{equation}
  R\equiv\int_0^\infty r\,dk = \int_0^\infty j(\phi)\,d\phi.\label{eq:def:R}
\end{equation}
The value of $R$, based on the numerical approximation to $j(t)$, is
\begin{equation*}
  R\sim 1.34.
\end{equation*}

%

Converting back to original physical units, the total density of
clusters produced during the nucleation era is
\begin{align}
\frac{R}{\nu}[r][k]&=\frac{R}{\nu}\brk{\frac{\Omega\eta_*}{d\Lambda^{2/3}\sig^2}e^{-g}}^\frac35,\label{eq:scales:density}
\intertext{where $g\equiv\frac{\sig^3}{2\eta_*^2}$ is the energy cost to create a critical cluster. Similarly, the characteristic cluster size $[k]$ and characteristic time of formation $[t]$ in original physical units are}
\Brk{k}&=\brk{\frac{d\eta_*^4}{\Omega\sig^3\Lambda}e^{g}}^\frac35,\label{eq:scales:k}\\
\Brk{t}&=\brk{\frac{\eta_*}{d\sig^2}}^\frac35 \brk{\frac{e^{g}}{\Omega\Lambda}}^{\frac25}.\label{eq:scales:t}
\end{align}

The nucleation asymptotics is not uniformly valid as $t\goto\infty$. 
The clusters grow by adsorbing monomers and the
super-saturation decreases according to \eqref{eq:conserv1}.
The nucleation era assumption that $\eta\sim1$ loses
validity.
In the next section, we assume a negligible amount of nucleation and
we follow the distribution of clusters as they evolve during the subsequent
growth era.

\section{The Growth Era}
The growth era begins when the nucleation of additional clusters becomes negligible.
The continued expansion of the existing clusters depletes the super-saturation, which in turn slows their further expansion.
The `tail' of the growth era is characterized by small super-saturation, $0<\eta\ll1$, and a narrow, almost steady distribution of cluster sizes.

Mathematically, the asymptotic solution to the SP during the growth era is specified by the size of the largest clusters, and the characteristic deviation from the largest size.
The largest cluster size and characteristic deviation satisfy simple ODE's, whose solution can be computed explicitly.

\subsection{New Variables}
We  introduce new variables for the location and width of the distribution. 
Let $K(t)$ be the size of the largest cluster as a function of time.
The support of $r(k,t)$ is  $0\le k\le K(t)$.
In this sense, $K(t)$ is the `front' of the distribution.
The largest cluster, like any other one, grows according to
$\dot K = u(K,\eta)$, where $u$ is taken from \eqref{eq:tag:s}.
It is convenient to introduce a `profile' variable, $x$, which represents translation and scaling of cluster size $k$:
\begin{equation}
  x \equiv  \frac{k-K(t)}{a(t)}.\label{eq:define:x}
\end{equation}
Heuristically, the scaling function $a(t)$ represents the characteristic deviation of cluster size $k$ from the front, $K(t)$.
Its precise meaning and determination are forthcoming.
The distribution of clusters in $x$-space, denoted by $q(x,t)$, is related to $r(k,t)$ by
\begin{equation}
 q(x,t) = \frac{a(t)}{R} r(K(t)+a(t)x,t).\label{eq:define:q}
\end{equation}
The front of the new distribution, $q(x,t)$, is located at $x=0$ and its support is in $x\le0$.
Recall that $R$ in \eqref{eq:define:q} is the integral of $r(k,t)$ defined in \eqref{eq:def:R}.
It is included in the definition of $q$ to normalize it so that
\begin{equation}
  \int_{-\infty}^0q\,dx=1.\label{eq:q:normal}
\end{equation}
Translating SP (\ref{eq:K:scaled:exact}--\ref{eq:conservation:kin}) for $r(k,t)$ into a SP for $q(x,t)$ yields
\begin{align} 
 \dot K &= K^\third\eta -s,\label{eq:K_dot_growth}\\
  0&=q_t+(wq)_x \text{ in } x<0, \label{eq:PDE_growth}\\
  w &= \frac{-\dot a\,x + (K+a\,x)^{\frac13}\eta -K^{\frac13}\eta}a,\label{eq:PDE_ad:vel}\\
  \eta &= 1-\frac{\eps^2R}{\sig^3}\int_{-\frac{K}{a}}^0 (K+a\,x)\, q\,dx.\label{eq:eta_relation}
\end{align}
Here, $q$ and $w$ are functions of $x$ and $t$, while $a,\ K$ and $\eta$ are
functions of $t$.
Notice that the transformed advection velocity in $x$-space, $w$, vanishes at $x=0$
regardless of $a(t)$. 
This guarantees that the `front' of the distribution remains at $x=0$.

We now turn to the precise determination of the characteristic deviation, or `width',  $a(t)$.
First, we can specify its time evolution so that the $x$-advection velocity, $w$, has $w_x(0,t)=0$, as well as $w(0,t)=0$.
Setting  $w_x(0,t)=0$ in \eqref{eq:PDE_ad:vel} gives the ODE 
\begin{align}
 \dot a &= \frac13 K^{-\frac23}a\eta.\label{eq:a:choice}
\intertext{Using this choice of $\dot a$, the advection velocity in \eqref{eq:PDE_ad:vel} becomes}
  w &= \frac{K^{\frac13}\eta}{a}\BRK{\brk{1+\frac{a}{K}\,x}^{\frac13} -\brk{1+\third\frac{a}{K}x}}, \label{eq:advection:1}
\end{align}
Assuming that $a\ll K$, the advection velocity $w$ in  \eqref{eq:advection:1} is $\O\brk{\frac{a}{K}\frac{\eta}{K^{2/3}}}$ for $x=\O(1)$.
We show that the asymptotic determinations of $a(t),\, K(t)$ and $\eta(t)$ indeed satisfy $a(t)\ll K(t)$ and $\eta\ll K^{2/3}$ during the growth era, and during the `beginning' of the coarsening era as well.
Hence, the $x$-distribution $q(x,t)$ is nearly time-independent. 
\subsection{Scalings (relative to nucleation era)}
\label{sec:growth:sub:scaling}
The super-saturation, $\eta$, goes from being close to $1$ (at the
tail of the nucleation era) to being close to zero (at the tail of the
growth era).
Therefore, we do not rescale $\eta$.
The dimensionless signaling problem (\ref{eq:K_dot_growth}, \ref{eq:PDE_growth}, \ref{eq:advection:1}, \ref{eq:eta_relation}, \ref{eq:a:choice}) is based on nucleation era units $[k]$ and $[t]$ of cluster size and time, in (\ref{eq:scales:k}, \ref{eq:scales:t}).
Hence rescaling of $K$ and $t$ based on dominant balance in the SP are relative to $[k]$ and $[t]$.
For instance, the dominant balance of terms in \eqref{eq:eta_relation} gives $\frac{\sig^3}{R\eps^2}$ as a `relative' unit of $K$, so the actual unit of $K$ is $[k]\frac{\sig^3}{R\eps^2}$.
Similarly, a dominant balance between $\dot K$ and $K^\third \eta$ in \eqref{eq:K_dot_growth} (neglecting the exponentially small $s$) gives the relative unit of time, $\frac{\sig^2}{\eps^{4/3}R^{2/3}}$, and the actual unit of time is $[t]\frac{\sig^2}{\eps^{4/3}R^{2/3}}$.
The natural unit of the advection velocity, $w$, is length over time. 
In our case, `length' is $x$ which has $O(1)$ units, and the units of
time are the units of $t$. 
Therefore, we use $\frac{\eps^{4/3}R^{2/3}}{\sig^2}$ as the units of $w$.
The ODE \eqref{eq:a:choice} is invariant under scaling of $a$. 
But, it follows from \eqref{eq:a:choice} and the leading approximation to \eqref{eq:K_dot_growth}, $\dot K\sim K^{1/3} \eta$, that $a$ is asymptotically proportional to $K^{1/3}$.
This suggests a unit for $a$,  $\frac{\sig}{R^{1/3}\eps^{2/3}}$.
The \emph{real} justification for the unit of $a$ comes from the
asymptotic matching between the growth and nucleation eras, where it is shown that the constant of proportionality between scaled $a$ and scaled $K^{1/3}$ is indeed a universal, dimensionless constant.

The scales are summarized in the table below.
\begin{center}
Scaling Table

\begin{tabular}{l*{6}{>{$}c<{$}}}
Variable&q&K&\eta&t&a&w\\
\hline\\[-4mm]
Unit&1&\tau^{-3}&1&\tau^{-2} &\tau^{-1}&\tau^2
\end{tabular}
\end{center}
Here, $\tau$ is defined as
\begin{equation*}
  \tau \equiv \frac{R^\third\eps^\frac23}{\sig}.
\end{equation*}
The scaled versions of  (\ref{eq:K_dot_growth},
\ref{eq:PDE_growth}, \ref{eq:advection:1}, \ref{eq:eta_relation}, \ref{eq:a:choice}) are
\begin{align}
  \dot K &= K^\frac13\eta-s\tau,\label{eq:K_ODE:1} \\
  q_t&=-(wq)_x,\label{eq:PDE:2}\\
   w &= \frac{K^{\frac13}\eta}{\tau^2a}\BRK{\brk{1+\tau^2\frac{a}{K}\,x}^{\frac13}
  -\brk{1+\frac{\tau^2}{3}\frac{a}{K}x}}\label{eq:advection:3}\\
  \eta &= 1-K\int_{-\frac{K}{\tau^2a}}^0\brk{1+\tau^2\frac{ax}{K}}q \,dx,\label{eq:eta:1}\\
  \dot a &= \frac13 K^{-\frac23}a\eta.\label{eq:ode:a:1}
\end{align}
\subsection{Reduced Growth Kinetics}
\begin{figure}[ht!]
  \centerline{
\parbox{9cm}{\centerline{
\resizebox{7cm}{!}{\includegraphics{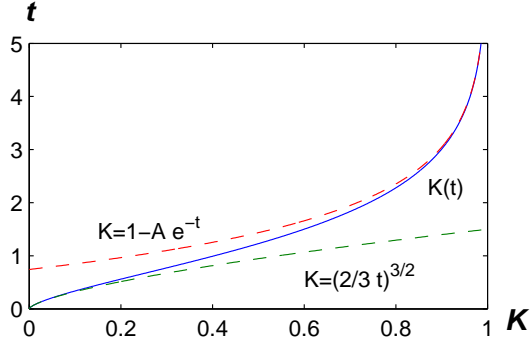}}}
\caption[The size of the largest cluster during the growth era]{The size of the largest cluster, $K(t)$, during the growth
  era. 
The two dashed lines show asymptotic behaviors of $K(t)$. 
For small values of
  $t$, \mbox{$K(t)\sim(\frac23t)^\frac32$} and  for large values of
  $t$, \mbox{$K(t)\sim 1-2.1e^{-t}$}.}
\label{fig:K_T}
}}
\end{figure}

The reduced equations that govern the dynamics during the growth era, are found by taking the $\eps\goto0$ limit of (\ref{eq:K_ODE:1}--\ref{eq:ode:a:1}) and using the normalization \eqref{eq:q:normal}.
The first four equations yield
\begin{align}
  \dot K &= K^\third\eta, \label{eq:growth:scaled}\\
  q_t&=0,\label{eq:q:indep}\\
  \eta &= 1-K. \label{eq:eta:growth:scaled}
\intertext{The advection velocity in \eqref{eq:advection:3} is
  $\O(\tau^2)$, and so it vanishes in the $\eps\goto0$ limit, as in
  \eqref{eq:q:indep}. 
We can therefore denote $q(x,t)$ as $Q(x)$ during the growth era.
  expected.
Equation \eqref{eq:ode:a:1} for $\dot a$ is invariant under the limit $\eps\goto0$. 
Using \eqref{eq:growth:scaled} we rewrite it as  $ 3\frac{\dot a}{a} =  \frac{\dot K}{K}$. This can be integrated, resulting in the aforementioned proportionality of $a$ to $K^{1/3}$.} 
  a &= C K^\third,\label{eq:a:1}
\end{align}
with the constant $C$ to be determined by asymptotic matching with the nucleation era.
We now determine $K(t)$. 
Substituting expression \eqref{eq:eta:growth:scaled} for $\eta$ into \eqref{eq:growth:scaled}
gives the
\begin{equation*}
  \dot K = K^\third(1-K). \label{eq:K:growth:ode}
\end{equation*}
The solution to \eqref{eq:K:growth:ode}, subject to $K(0)=0$, is given
  implicitly by\footnote{The trivial
  solution where $K(t)=0$ is ignored; we assume that $K(t)\ne0$ for $t>0$.}
\begin{equation}
  t = \sum_{j=0}^2 r_j \log\brk{1+r_j K^\frac13}.
\label{eq:K:impl:sol}
\end{equation}
Here, $r_j$ are the cubic roots of $-1$: $r_j=e^{\frac{2j+1}{3}i\pi}$.
Figure~\ref{fig:K_T} shows the solution together with its asymptotic behavior as $t\goto 0$ and $t\goto\infty$.


\subsection{Asymptotic Matching with the Nucleation Era}
\begin{figure}[ht]
  \centerline{
\parbox{8cm}{\centerline{
\resizebox{7cm}{!}{\includegraphics{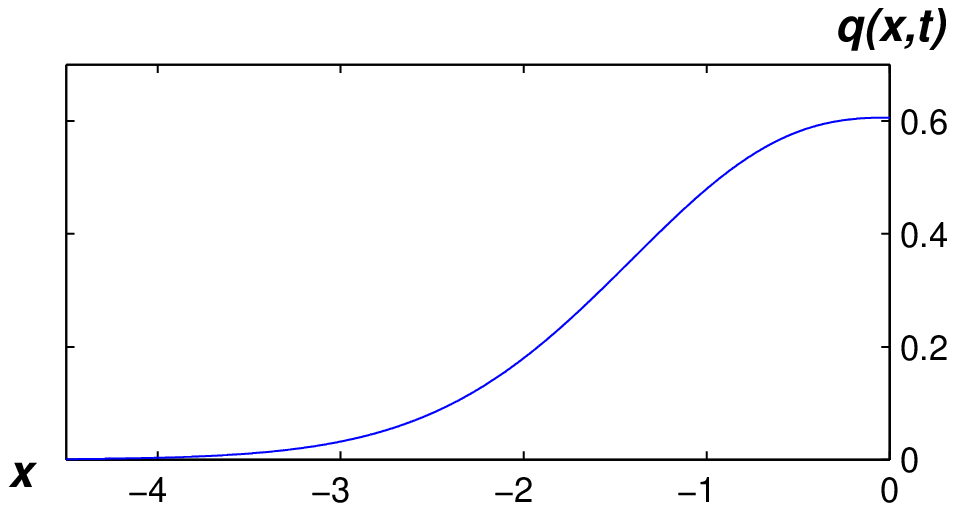}}}
\caption[The cluster size distribution during the growth era]{The cluster size distribution \mbox{$q(x)=\frac{C}{R}j(-Cx)$}, with $C\sim0.81$.}
\label{fig:distrib_prof}
}}
\end{figure}

We determine $C$ and $Q(x)$ by asymptotic matching of the nucleation
and growth era solutions. 
This is done by examining the nucleation era solution for time that is long relative to the nucleation time  and short relative to growth time, that is $\tau^2\ll t \ll 1.$
During the nucleation era, the density of clusters is given by \eqref{eq:char:sol}
\begin{equation}
  r(k,t) = k^{-\third}j\fbrk{t-\frac32k^\frac23}, \qquad j(t)=e^{\delta\eta(t)}.\label{eq:nucleation:2}
\end{equation}
Here, $t$ and $k$ have \emph{nucleation} era scalings.
Writing \eqref{eq:nucleation:2} in the growth era variables using
\eqref{eq:define:x} and \eqref{eq:define:q}, and giving $t,\, K$ and $a$  growth era scaling, we get
\begin{equation}
  q(x,t) =\frac{a}{\tau R}\,\brk{\frac{K}{\tau^3}+\frac{ax}{\tau}}^{-\third} j\fbrk{\frac{t}{\tau^2}-\frac32\brk{\frac{K}{\tau^3}+\frac{ax}{\tau}}^\frac23},\label{eq:r:as:q:growth}
\end{equation}
The asymptotic behavior of $K(t)$ is taken from the  $t\goto0$ of the
\emph{growth} era,
\begin{equation}
  K(t) \sim \brk{\frac23t}^\frac32, \text{ for } t\ll 1. \label{eq:K:small:time}
\end{equation}
In appendix \ref{sec:nuc:grow:match} we show that substituting
\eqref{eq:K:small:time} into \eqref{eq:r:as:q:growth} results in
\begin{equation}
   q(x,t)=Q(x) = \frac{C}{R} j\fbrk{-Cx}, \text{ for } \tau^2 \ll t \ll 1.\label{eq:match:NG}
\end{equation}
Examining \eqref{eq:match:NG} we see that the long-time limit of the
nucleation era solution, written in the variables $q$ and $x$, is time-independent, agreeing with \eqref{eq:q:indep}.
The asymptotic matching does not uniquely determine $q$ and $C$: different choices of $C$ lead to differently scaled distribution functions $q$.
We are free to  normalize $q$ by choosing the constant $C$ so that
\begin{equation*}
   \int_{-\infty}^0\hspace{-1ex} x\,Q(x)\,dx=-1.
\end{equation*}
This determines a unique value for $C$,
\begin{equation*}
   C=R\int_0^\infty \hspace{-1ex}t\,j(t)\,dt \sim 0.81.
\end{equation*}
The value $0.81$ of $C$ is based on the numerical solution of the integral equation \eqref{eq:integral_1}.
Figure~\ref{fig:distrib_prof} shows the normalized distribution
profile, $Q(x)$.

While the choice of constant $C$ is arbitrary, it only affects $q$ and not the solution in `real' cluster size, $k$. 
Reverting back to $r(k,t)$, we find that the distribution of cluster
sizes during the growth era is given by
\begin{equation}
  r(k,t) =\frac{1}{\tau^{-1}K^\third}j\brk{\tau^{-2}\frac{K-k}{K^\third}},\label{eq:r:growth}
\end{equation}
which is independent of our choice of constant $C$. 
In \eqref{eq:r:growth}, the time $t$ is measure using growth era
scales, $\tau^{-2}$, and $k$ is measured using the same scales as $K$, $\tau^{-3}$.

\subsection{Physical Predictions}
The characteristic time, $[t]_{\text{growth}}$, of the growth era in original
physical units is:
\begin{align}
[t]_{\text{growth}}&\sim\brk{\frac{\eta_*}{d\sig^2}}^\frac35
\brk{\frac{e^{g}}{\Omega\Lambda}}^{\frac25}\frac{\sig^2}{R^{2/3}\eta_*^{4/3}}.
\intertext{Recall $g\equiv\frac{\sig^2}{2\eta_*^2} $ is the energy
  cost to create a critical cluster.
The characteristic cluster size is,}
[k]_{\text{growth}}&\sim \brk{\frac{d\eta_*^4}{\Omega\sig^3\Lambda}e^{g}}^\frac35 \frac{\sig^3}{R\eta_*^2}.
\end{align}
Since the distribution's width $a\sim\tau^{-1}C$ is much smaller
  than its location, $K\sim\tau^{-3}$, the
clusters all have approximately the same size, $[k]_{\text{growth}}$.

\section{Coarsening Era}
Within the growth era approximation, the cluster distribution has a definite limit as $t\goto\infty$. 
In this limit the distribution is concentrated in an $\O(\tau^{-1})$
interval at $k\sim\tau^{-3}$ (recall that
$\tau=\frac{R^{1/3}\eps^{2/3}}{\sig}$), and the super-saturation
vanishes.
 
This is not the whole story: the growth era approximation neglects the surface energy component, $s$, of the advection velocity in \eqref{eq:K_ODE:1}.
The remaining component of the advection velocity is proportional to
the super-saturation, which then depletes to zero as the clusters grow.
The full advection velocity for $K$, which does not neglect $s$,
\eqref{eq:K_dot_growth} has a fixed point at a slightly smaller value
of $K$ than predicted in the growth era.
A smaller value of $K$ implies that the super-saturation maintains a
small, yet non-zero value in this limit, which, in turn, implies that
the width of the distribution, $a$, will continue to grow, according
to \eqref{eq:ode:a:1}, with a characteristic time which sets the
timescale of the the era. 
The widening of the initially narrow distribution is only a precursor
to the era. 
Once the distribution is not narrow, the effective advection velocity, $w$ in \eqref{eq:advection:3} is no longer negligible and the distribution profile changes. 
The distribution eventually asymptotes to a self-similar solution of
the original LS theory.

In this section, we scale the equations and find a simple advection
PDE that captures the dynamics of the solution during the coarsening era. 
The initial conditions for this PDE are found by matching the
$t\goto0$ limit of the coarsening era solution with the $t\goto\infty$
limit of the growth era. 
Using the initial conditions found by this matching process we evolve
the coarsening era solution using a numerical solver. 

\subsection{New Variables and Scaling }
The coarsening era is characterized by the balance between the
 terms in the advection velocity \eqref{eq:tag:s} and a small
 super-saturation. 
The distribution of cluster sizes slowly widens.
Eventually, it fills the whole interval $\Brk{0,K}$.
Thus, the description of the distribution as an advecting narrow
 profile is not valid here, and the subsequent choice for the `width'
 of the distribution, $a$ in \eqref{eq:a:choice} is not convenient for
 this era.
Instead of concentrating on a narrow distribution near the largest
 clusters, we consider a distribution that will be as wide as
 $K$.
Thus in (\ref{eq:K_dot_growth}--\ref{eq:eta_relation}) we now chose $a$ so:
\begin{equation}
  a(t) = K(t).
\end{equation}
Consequently, $x$ as in \eqref{eq:define:x} is no longer a
useful variable for the distribution. 
Instead we define
\begin{equation}
  y \equiv x+1, \quad q(y,t) = \frac{K}{R}r(Ky,t). \label{eq:y_and_q:def}
\end{equation}
This is equivalent to defining $y = \frac{k}{K}$, as can be seen from \eqref{eq:define:x}.
Using this choice for $a$ and the variable $y$ in equations
(\ref{eq:K_dot_growth}--\ref{eq:eta_relation}) gives the following
equations, written using the nucleation scaling.
\begin{align}
  0&=q_t+(wq)_y, \text{ for } 0\le y\le1,\label{eq:advection:precoarse}\\
  w &=\eta K^{-\frac23}\brk{ y^\third-y}+K^{-1}s(y-1),\label{eq:advection:precoarsen}\\
\eta&=1-\tau^3K\int_0^1 y\, q\,dy,\label{eq:eta:precoarsening:1}\\
 \dot K &= K^\third\eta -s.\label{eq:K:dot:precoarse}
\end{align}

As in the growth era, the conservation equation
\eqref{eq:eta:precoarsening:1}  implies a scale of $\tau^{-3}$ for $K$.
The scale of the super-saturation, $\eta$ is found from the dominant
balance of the terms in the RHS of the equation for $\dot K$
\eqref{eq:K:dot:precoarse}, which implies $\eta = \tau s$
The time-scale is much larger than that of the growth era.
Dominant balance of all three terms in \eqref{eq:K:dot:precoarse}
gives the characteristic time $\oneover{s\tau^3}$.
Similarly, either term in the RHS of \eqref{eq:advection:precoarsen} implies a scale $s\tau^3$ for $w$.

The scales of the variables are summarized in the following table. 
\begin{center}
Scaling Table\\
\begin{tabular}{l*{5}{>{$}c<{$}}}
Variable& \eta&K&t&w&q\\
\hline\\[-4mm]
Unit&s\tau&\tau^{-3}&\oneover{s\tau^3}&s\tau^3&1
\end{tabular}
\end{center}
These scales are relative to those of the nucleation era. 

\subsection{Reduced Coarsening Kinetics}
We rewrite equations (\ref{eq:advection:precoarse}--\ref{eq:K:dot:precoarse})  using the scales above.
In the $\eps\goto0$ limit they read
\begin{align}
  0&=q_t+(wq)_y, \text{ for } 0\le y\le1,\label{eq:advection:coarse}\\
w &= K^{-\frac23}\eta\brk{y^\third-y}+K^{-1}(y-1),\label{eq:advection:coarsening:scaled}\\
0 &= 1-K\,M,\label{eq:eta:delicate}\\
\dot K&=K^\frac13\eta-1.\label{eq:K:coarse}
\intertext{Here, $M$ is the first moment of $q$,}
M&=\int_0^1y\,q\,dy,\label{eq:def:M}
\end{align}
itself a function of $t$.
The value of the supersaturation, $\eta$, is \emph{implicit} in this
system of equations. 
In the growth era, it is given explicitly by equation
\eqref{eq:eta:growth:scaled}, however, in the coarsening era, the
scales have conspired so that $\eta$ vanishes from the corresponding equation
\eqref{eq:eta:delicate}.
An explicit expression for the value of $\eta$ can be
extracted from the other equations.
The details of this derivation are in appendix
\ref{sub:sec:finding:eta}\@.
Here, we give the results.

During the coarsening era, the super-saturation, $\eta$ is given by
\begin{equation}
  \eta = \frac{M^\third M_0}{M_\third}\label{eq:eta:simple},
\end{equation}
where, $M_0$ and $M_\third$ are moments of the distribution $q$,
\begin{equation}
  M_0 = \int_0^1 q\,dy,\qquad  M_\third = \int_0^1 y^\third q\,dy,\label{eq:def:M_0}
\end{equation}
and $M$ is defined in \eqref{eq:def:M}.
Substituting \eqref{eq:eta:simple} into  the PDE  (\ref{eq:advection:coarse}, \ref{eq:advection:coarsening:scaled}) for $q$ and replacing $K$ by $\oneover{M}$ using \eqref{eq:eta:delicate}, we find 
\begin{align}
  0&=q_t+(wq)_y, \quad 0\le y\le1,\label{eq:PDE:coarsen}\\
  w &= M\brk{\frac{M_0}{M_{\third}}\brk{y^\third-y}+(y-1)}.
\label{eq:advection:coarsen}
\end{align}
This PDE is nonlinear due to the dependence of the advection velocity
on the moments of the solution. 
The distribution that emerges from the growth era is very narrow and a linear approximation of the advection velocity \eqref{eq:advection:coarsen} results in a PDE that can be solved analytically. 
The analytic solution widens and eventually the linear approximation loses validity and we turn to a numerical solution. 
The PDE (\ref{eq:PDE:coarsen}, \ref{eq:advection:coarsen} has similarity solutions and the numerical solution converges to one of them as $t\goto\infty$. 
\subsection{Asymptotic Matching with the Growth Era}
The distribution of cluster sizes that
emerges from the long time limit of the growth era, is concentrated in
a narrow distribution near $y=1$. 
The distribution's width is $\O(\tau^2)$ and thus, not uniformly valid
as $\eps, \tau\goto0$: A `point' distribution  at $y=1$ is stationary
under the action of the PDE, but a distribution with any finite width
widens.
To find effective initial conditions, which are
valid as $\eps, \tau\goto0$, we asymptotically match the coarsening
era solution to the long-time limit of growth era.
Initially, we approximate the solution by a narrow distribution slowly widening due to the linearization of the PDE.
Eventually, non-linear effects must not be ignored and this approximation loses validity.
At this point we turn to a numerical solver to follow the continued evolution of the solution.

The long time limit of the growth era is found from \eqref{eq:r:growth}.
Taking the limit $K\goto1$ we find
\begin{equation*}
r(k,t) = \oneover{\tau^2 R}j(\tau^{-2}-\tau k)
\end{equation*}
Na\"ively, we expect this to provide the IC for $q(y,0)$. 
We pursue this venue only to change it slightly later.
Thus, from the definition of $q(x,t)$ in \eqref{eq:define:q} we have
\begin{equation}
q(y,0)= \oneover{\tau^2R} j\brk{\frac{1-y}{\tau^2}}.\label{eq:q:init:coarsen}
\end{equation}
and 
\begin{equation*}
K=M=M_0=M_{\third}=\eta(0)=0.
\end{equation*}
As we see, in $y$-space, this is a very narrow distribution and we approximate PDE (\ref{eq:PDE:coarsen}--\ref{eq:advection:coarsen}) by its leading order term about $y=1$
\begin{align}
 q_t + (w q)_y =0,\\
 w = \third (y-1).
\end{align}
The solution to this PDE subject to the IC in \eqref{eq:q:init:coarsen} is
\begin{equation*}
q(y,t)= \oneover{e^{(t-t_0)/3} R} j\brk{\frac{1-y}{e^{(t-t_0)/3}}},\quad t_0\equiv-6 \log \tau.
\end{equation*}
This widening solution is valid as long as it it narrow, that is, as
long as
\begin{equation*}
  e^{(t-t_0)/3} \ll 1.
\end{equation*}

We see that the na\"ive asymptotic matching leads to an initial
condition that depends on $\tau$ and therefore, cannot be used
directly as a universal solution $q(y,t)$ that is valid in
the coarsening era regardless of $\tau$. 
However, by shifting the origin of the coarsening time, by $t_0$, the
different functions $q(y,t)$ for different values of $\tau$, collapse
onto the same function:
\begin{equation*}
q(y,t+t_0)= \oneover{e^{t/3} R} j\brk{\frac{1-y}{e^{t/3}}}.
\end{equation*}
The timeshift $t_0$ is dependent on $\tau$, but the resulting $q$ is
not.
To summarize: we shift time in the coarsening era so that the
asymptotic matching with the growth era happens at $t=-t_0=6\log \tau$. 
This results in a universal function $q(y,t)$ that captures the
distribution on cluster sizes in $y$ space during the coarsening era,
and is uniformly valid for all $\tau\ll 1$.
During an initial phase, the distribution is narrow, and  an analytic approximation
for $q$ is given by
\begin{equation}
q(y,t)= \oneover{e^{t/3} R} j\brk{\frac{1-y}{e^{t/3}}}. \label{eq:sol:precoarse}
\end{equation}
At $t=0$ the distribution $q$ so wide that the linear approximation
for the PDE must not be trusted. 
We therefore, must take the solution $q$ from some time $t<0$ when the
distribution is still narrow, and use it as IC for a numerical solver
that uses the full, non-linear PDE to approximate $q(y,t)$ for larger
values of $t$.

\subsection{Numerical Solution}
For the solution of the full PDE
(\ref{eq:PDE:coarsen}, \ref{eq:advection:coarsen}) we turn to a
numerical solver.
We assume that the numerical solver can accurately
resolve distributions that are wider than a finite size $\delta$.
As we have seen, the distribution, $q(y,t)$,
starts  with width $\tau^2$ (presumably with $\tau^2<\delta$) at
$t=6\log \tau$, and widens according to \eqref{eq:sol:precoarse}. 
As the  distribution $q(y,t)$ widens, the linear approximation to the
advection velocity loses validity.
Thus, $\delta$ must be small enough so the linear approximation to the
PDE is still valid when the distribution has width $\delta$.

The solution has width $\delta$ when the shifted time, $t$, satisfies
$e^{t/3}=\delta$. 
In other words, we start the numerical solver with the IC in
\eqref{eq:sol:precoarse} with 
\begin{equation*}
  t = 3\log \delta.
\end{equation*}
The resulting solution is globally valid. 
Different values of $\tau$ will imply a longer time during which the
distribution is widening, \emph{prior} to the numerical solution. 
In the resulting distribution $q(y,t)$ the origin of time must be
shifted back by $t \goto t + t_0 = t - 6\log \tau$.


\begin{figure}[ht!]

 \centerline{
\parbox{11cm}{\centerline{
\resizebox{7cm}{!}{\includegraphics{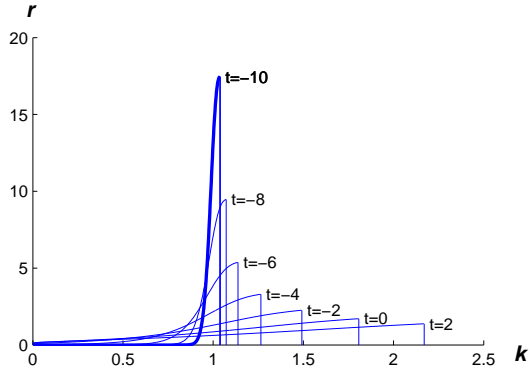}}}
\caption[The cluster size distribution during the coarsening era]{The numerical solution at various times as found using
  clawpack. The dark curve is the initial condition, found from 
  \eqref{eq:sol:precoarse} at $t=-10$. The horizontal axis is $k$ (in units of
  $\tau^{-3})$. The solution continues to evolve after $t=0$, but the
  dominant change is a linear stretching of the distribution.}
\label{fig:coarsening_10}
}} 
\end{figure}

We acquired numerical results by using LeVeque's 
conservation law PDEs software package, \verb!clawpack! \cite{LeVeque}, with
the Riemann solver \verb!rp1adecon! (a Riemann solver for conservative advection).
To find a numerical solution using \verb!clawpack!, we used a grid
size of $1/500$ for the numerical
solver, and chose a value
of $e^{-10/3}\sim 0.0357$ for $\delta$.
This corresponds to starting the numerical solver at $t=-10$.
The moments, $M,\ M_0$, and $M_\third$ are computed by the trapezoidal rule.
By comparing to numerical results obtained from a smaller $\delta$ and grid
spacing we estimate the error to be $\sim 1\%$.

To reconstruct $r(k,t)$ we inverted the relation for $q(y,t)$ in
\eqref{eq:y_and_q:def}  using $K$,  found from (\ref{eq:eta:delicate},
\ref{eq:def:M}).
This results in a distribution in $k$-space, with $k$ and $t$ still
scaled with the coarsening era scales, $tau^{-3}$ and $\oneover{s\tau}$.
\begin{equation*}
  r(k,t+t_0) = RMq(kM,t), \quad t_0 = -6\log \tau.
\end{equation*}
Figure~\ref{fig:coarsening_10} shows  the solution, $r(k,t)$, at various values of $t$. 

Due to our definition of the origin of coarsening time, $t=0$, the interesting
part of the solution happens at $-10<t<2$. 
The solution at times $6\log \tau<t<-10$ is captured by the
widening solution \eqref{eq:sol:precoarse}.
At large $t$, the numerical solution asymptotes to a similarity
solution of the PDE (\ref{eq:PDE:coarsen},
\ref{eq:advection:coarsen}). 
In the next section, we review the derivation of the family of
similarity solutions and identify among it the long-time limit of the
numerical solution.
\section{Similarity  Solutions}

The coarsening era equations (\ref{eq:def:M}, \ref{eq:def:M_0}--\ref{eq:advection:coarsen})  have a family of
similarity solutions, as shown by LS.
Here, we review the derivation and show that the family of similarity
solution can be parametrized by their order of contact with zero at
the largest clusters. 
This allows us to predict \emph{a priori} the self-similar solution to
which the numerical solution will converge. 
In the end of the section, this prediction is confirmed by comparing
the numerical solution to the predicted similarity solution.

To find the family of similarity solutions we look for a solution $q(y,t)$ of the
PDE (\ref{eq:PDE:coarsen}, \ref{eq:advection:coarsen}) using
separation of variables into  `temporal' and  `spatial` functions,
$c(t)$ and $P(y)$:
\begin{equation}
q(y,t)=c(t)\,P(y).\label{eq:ss:ansatz}
\end{equation}
First, we find the temporal part, $c(t)$.
Recall the definition of $M$:
\begin{equation*}
  M=\int_0^1 y\,q(y,t) \, dy=c(t)\int_0^1  y\,P(y)\, dy.
\end{equation*}
Hence, for a similarity solution as in \eqref{eq:ss:ansatz}, $M$ is proportional to $c$. 
We denote the constant of proportionality by $F$:
\begin{equation}
  F\equiv \int_0^1 y\,P\, dy.
\end{equation}
In other words we have that,
\begin{equation}
  M=c\,F \quad\Rightarrow\quad \dot M=\dot cF. \label{eq:M:dot:1}
\end{equation}
Another equation for $\dot M$ can be derived from the definition of
$M$, the PDE (\ref{eq:PDE:coarsen}) and integration by parts,
\begin{equation}
  \dot M  = \int_0^1 y\, q_t =- \int_0^1y (wq)_y \, dy = \int_0^1  wq \, dy.\label{eq:M:dot:2}
\end{equation}
From the two equations \eqref{eq:M:dot:1} and \eqref{eq:M:dot:2} for
$\dot M$, we derive an ODE for $c$:
\begin{equation*}
  \dot c F = M^2\brk{1- \mu}.
\end{equation*}
Here, $\mu$, is 
\begin{equation*}
  \mu \equiv
  \frac{M_0}{M_\third}=\frac{\int_0^1P\,dy}{\int_0^1yP\,dy}.
\end{equation*}
Thus, $c$ satisfies the ODE
\begin{equation}
  \dot c = -Fc^2(\mu-1).\label{eq:c_dot}
\end{equation}
Notice that $0\le\mu\le1$ and that $\mu$ is time-independent, hence the solution of ODE \eqref{eq:c_dot} is 
\begin{equation}
  c(t) = \oneover{F(\mu-1)(t-t_s)}\label{eq:c_of_t},
\end{equation}
for some $t_s$.

Now that we have $c(t)$, we can find the spatial part of the
similarity solution, $P(y)$.
Substituting the \eqref{eq:ss:ansatz}  into the advection PDE
\eqref{eq:PDE:coarsen}, and using the ODE \eqref{eq:c_dot} for $c$, yields an ODE for P:
\begin{equation}
  P\brk{c^2F(1-\mu) + c w_y} + P_y  c w=0 \quad \Rightarrow \quad  \frac{P_y}{P}= -\frac{M(1-\mu) +  w_y}{w}.\label{eq:ODE:P}
\end{equation}


\begin{figure}[ht!]
  \centerline{
\parbox{8cm}{\centerline{
\resizebox{6cm}{!}{\includegraphics{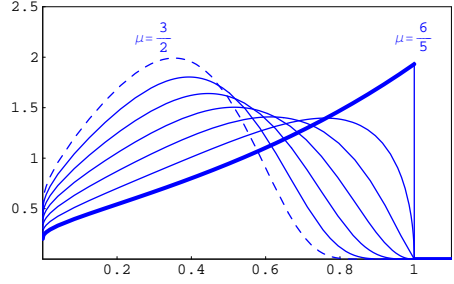}}}
\caption[Various similarity solutions]{The plot of the similarity solution for various values of
  $\mu$. The solutions are normalized so that $M_0=1$. The dark line
  has $\mu=\frac65$.}
\label{fig:similarity:sol}
}}
\end{figure}
The explicit solution of \eqref{eq:ODE:P}, subject to $P(0)=1$, is
\begin{equation}
  P(x) = \exp\brk{\int_0^x \frac{\mu(2-\third y^{-\frac23})-2}{\mu(y^\third-y)+(y-1)}dx}.\label{eq:SS:integrated}
\end{equation}
The product of $c(t)$ and $P(y)$ gives a two-dimensional family of
solutions parametrized by $\mu$ and $t_s$.
The parametrization by $t_s$ in \eqref{eq:c_of_t} reflects
time-independence of the PDE (\ref{eq:PDE:coarsen},
\ref{eq:advection:coarsen}).
Different values of $\mu$ select different
similarity solutions for  $\mu\in\Brk{0,1}$.
However, we only allow values $\frac56\le\mu\le\frac32$.
Other values of $\mu$ generate $P$ that explodes at one of the
end-points, and we dismiss such solutions as non-physical.
The parametrization by $t_s$ in \eqref{eq:c_of_t} reflects
time-independence of the PDE (\ref{eq:PDE:coarsen},
\ref{eq:advection:coarsen}).
We use this symmetry during the asymptotic matching to the numerical solution.
Figure~\ref{fig:similarity:sol} shows a few similarity solutions for different values of $\mu$.
\subsection{Order of Contact at $y=1$}
Here, we find which of the similarity solution profiles in
\eqref{eq:SS:integrated} is the limit of the numerical solution in the coarsening era.
For this we study the order of at $y=1$ contact of the various
similarity solutions. 
The similarity solutions are originally parametrized by $\mu$.
It it more instructive to parameterize them by their order of contact
with zero.
At the tail-end of the coarsening era, the numerical solution
 is  discontinuous at $y=1$.
This corresponds to an order of contact 0. 
Thus, we find the similarity solution with order of contact zero,
and confirm that the numerical solution converges to it.

The order of contact is the power $p$ so that 
\begin{equation*}
  P(y)\sim b \,(1-y)^p \text{ as } y\goto 1^-,
\end{equation*}
for some constant $b\ne0$.
A little algebra shows that 
\begin{equation}
  p=\lim_{y\goto1^-} (\ln P)_y (1-y) =\lim_{y\goto1^-} \frac{P_y (y-1)}{P}.\label{eq:order:p}
\end{equation}
Substituting \eqref{eq:ODE:P} into \eqref{eq:order:p} and using
l'H\^opital's rule we find that the order of contact of the similarity
solution $P(y)$ is
\begin{equation}
  p=\frac{5\mu-6}{2\mu-3}.\label{eq:contact}
\end{equation}
Thus, we expect the numerical solution of the coarsening era to
converge to the similarity solution with $p=0$, hence
\begin{equation}
  \mu=\frac65.
\end{equation}
Equation \eqref{eq:ODE:P} is  integrable for this value of $\mu$,
resulting in 
\begin{equation}
P(y)= \frac{125 \exp\brk{-2 \sqrt{\frac{3}{7}} \left(\coth ^{-1}\left(\sqrt{21}\right)-\tanh ^{-1}\left(\frac{2
   \sqrt[3]{x}+1}{\sqrt{21}}\right)\right)}}{\left(5-x^{2/3}-\sqrt[3]{x}\right)^3}.
\end{equation}
Here, $P(y)$ is normalized so that
\begin{equation}
  \int_0^1 P \,dy=1,\qquad \int_0^1 y P \,dy \sim 0.632573, \qquad \int_0^1.
  y^\third P \,dy = \frac56.
\end{equation}
The dark, discontinuous line in Figure~\ref{fig:similarity:sol} shows
$P$ with  $\mu=\frac65$. 

\subsection{Asymptotic Matching with Coarsening Era}
%
%
\begin{figure}[ht!]
  \centerline{
\parbox{8cm}{\centerline{
\resizebox{6cm}{!}{\includegraphics{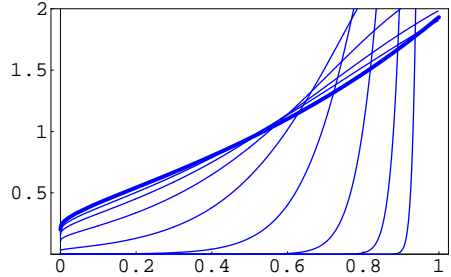}}}
\caption[The numerical solution approaching a similarity solution]{The numerical solution $q(y,t)$, scaled so that $M=1$, at
  different times. Starting on the right at $t=-10$ in a narrow distribution, and converging to the similarity solution with $\mu=\frac65$ (Dark
  line) at $t=4$.}
\label{fig:converge}
}}
\end{figure}

As can be seen in Figure~\ref{fig:converge}, the numerical solution
of the coarsening era converges to the similarity
solution with $\mu=\frac65$ as $t\goto\infty$. 
The temporal behavior of the similarity solution given by
\eqref{eq:c_of_t} contains an undetermined constant, $t_s$.
By asymptotic matching  of the numerical solution to the
similarity solution we now determine this last undetermined constant.

From \eqref{eq:M:dot:1} for $M$ and \eqref{eq:c_of_t} for $c(t)$ it
follows that the first moment of the similarity solution decays like
\begin{equation}
  M(t) = \frac{1}{(\mu-1)(t-t_s)},
\end{equation}
for some $t_s$.
Hence, a numerical solution that converges to the
similarity solution with $\mu=\frac65$,  must satisfy
\begin{equation}
  \frac5{M(t)} = t-t_s,\quad \text{ as } t\goto \infty.
\end{equation}

\begin{figure}[ht!]
\centerline{
\parbox{8cm}{\centerline{
\resizebox{6cm}{!}{\includegraphics{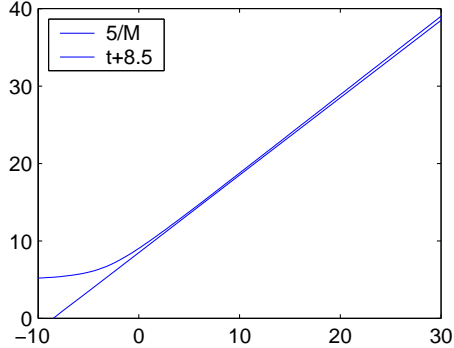}}}
\caption[Finding the time-shift for the similarity solution]{The match of $\frac{\mu-1}{M}$ to $t-t_s$. The
  approximate numerical value of $t_s$ is -8.5.}
\label{fig:finding_t_s}
}}
\end{figure}

From the numerical solution we calculate $M(t)$ and plot
$\frac{5}{M(t)}$.
By fitting a line of slope 1 to the linear part of the plot
of $\frac5M$ we find that $t_s\approx -8.5$.
The results are plotted in figure \ref{fig:finding_t_s}.

Figure~\ref{fig:finding_t_s} shows that the long-term behavior
of the numerical solution's first moment matches the predicted decay
as $t\goto\infty$. 
The same can be said about the profile of the numerical solution. 
Figure~\ref{fig:converge} show the convergence of the (scaled)
numerical solution to the similarity solution with $\mu=\frac65$.
The figure seems to indicate that at $t>5$ the numerical solution
should be close to the similarity solution. 
Indeed, at $t=5$ the numerical solution differs from the $\mu=\frac65$
similarity solution by less than $1\%$.
\subsection{Physical Predictions}
Our asymptotic solution to the LS PDE  predicts the emergence of
a distribution of clusters which is \emph{discontinuous} at the large
clusters. 
The predicted similarity solution, has a specified timeshift $t_s$. 
This does not include the timeshift $t_0$ which is needed for the
initially narrow distribution to widen prior to the numerical
solution. 
The scaling gives us specific physical scales for the
size of the clusters, $[k]_c$, and the time of their formation, $[t]_c$.

\begin{align}
[t]_c&\sim\brk{\frac{\eta_*^4d e^g}{\sig^3\Lambda\Omega}}^\frac35
\brk{R\eta_*^2}^{-\third}
 ,\\
[k]_c&\sim \brk{\frac{d\eta_*^4}{\Omega\sig^3\Lambda}e^{g}}^\frac35 \frac{\sig^3}{R\eta_*^2}.
\end{align}
\section{Conclusions}

The three eras, nucleation, growth, and coarsening emerge from
the asymptotic analysis of the aggregation model which combines
elements from BD and LS models. 
At the small super-saturation limit the three eras are separated by
increasingly longer times scales. 
The solution starts with IC $r(k,0)\equiv 0$, corresponding to pure
monomer. 
Each era provides effective initial conditions for the next one by
means of asymptotic matching, and the coarsening era selects the discontinuous
similarity solution as the global long0time limit of the aggregation
process.
The selection of the similarity solution is accompanied by physical
scales and a specific shift $t_s$ in \eqref{eq:c_of_t}, both were
previously unknown.
Due to the vast difference in scales between the
different eras, we do not expect a single experiment to resolve the
complete nucleation process, from nucleation to coarsening. 
However, different experiment could be set-up to examine each of the
different eras by adjusting the physical parameters.

Future work is possible in several aspects of this solution:

\begin{enumerate}
\item{ Post-coarsening effect of the ``ignition'' and diffusion}

The ignition phase, during which the nucleation rises from zero to the
rate specified by the Zeldovich formula,  takes a small, yet finite
time.
This implies that the front is not sharp, but rather transitions
smoothly to zero.
In addition, in deriving the advection PDE a diffusion term was
ignored because it is asymptotically small at all 3 eras.
Both effects imply that the order of contact with zero
at the front is not 0, and that there is be another scale of $k$ in which
the distribution transitions smoothly to zero near $k\sim K$. 
Therefore, at a time-scale even larger than that of the coarsening
era, this `contact layer' can widen  and possibly change the selection
 of the long-term similarity solution. 
Whether or not this is \emph{physically} relevant depends on the
  timescale and in which this happens. 
\item{ A successor to the Zeldovich nucleation rate}

The Zeldovich nucleation rate is calculated using a mean-field
  for the monomer distribution around a cluster. 
This approach does not take into account the role
  of fluctuation in the monomer density.
Aggregation is a discrete process, and nucleation a rare-event; the
  nucleation rate might be found by taking this into account.
\item{ Corroboration with experiments}

The scales predicted in the present chapter have not been checked against data
from experiment. 
The main difficulty in doing this is that in the experiment the large clusters tend
to fall out of the solution due to gravity. 
There have been some experiments in micro-gravity and it would be
informative to compare our results to the experiments. 
\item{ Extension to two dimensional models}

The present thesis models space and clusters as 3-dimensional.
This assumption sets the exponents governing the advection velocity of
the PDE\@. 
Two-dimensional crystals can be grown on silicon wafers and other flat
surfaces. Other experiments show growth of three dimensional crystals
on silicon wafers.
Crystals growing on a two-dimensional substrate will form large
clusters without gravity interfering and will allow for long experiments. 
Reformulating the current model for other dimensions and comparing
with experimental data could be fruitful.

\item { Inhomogeneous far-field supersaturation}

Our model assumes that the \emph{far-field} supersaturation is
constant. This has been shown to be an unstable solution. 
A model which allows for the far-field supersaturation to vary in
space will present a more realistic solution. 
\end{enumerate}

\bibliographystyle{amsalpha}
\bibliography{general}

\appendix{}
\chapter{Appendix}

%
%
%
\section{Deterministic Cluster Growth According to BD}
\label{sub:sec:det:BD}
Here we determine the domain of $k$ in which the average change
in the cluster size $\Average{\delta k}$ is a good approximation to the actual
change in cluster size. 
BD assumes that the cluster grows and shrinks by means of discrete,
independent Poisson processes. 
The cluster gains monomers at a rate $c_k\rho_1$ via the adsorption process, and via the emission process it loses monomers with rate
$d_k$. 
Therefore, the average change in size over a short interval $\delta t$ is 
\begin{equation*}
  \Average{\delta k} = \Average{k(t+\delta t)-k(t)} = (c_k\rho_1-d_k)\delta t,
\end{equation*}
provided that $\delta t$ is small enough so that 
\begin{equation}
\Abs{c_k\rho_1-d_k}\delta t \ll k.\label{eq:delta:k}
\end{equation}

Since the two controlling process are assumed to be independent
Poisson processes, the mean square deviation of $k$ from the average
is 
\begin{equation*}
  \Average{(\delta k -\Average{\delta k})^2} =  (c_k\rho_1+d_k)\delta t
\end{equation*}
The process can be considered deterministic if the deviation is much
smaller than the expected change, so
\begin{equation}
    \sqrt{c_k\rho_1+d_k}\sqrt{\delta t}\ll \Abs{(c_k\rho_1-d_k)\delta t}.\label{eq:deterministic}
\end{equation}
Combining inequalities \eqref{eq:delta:k} and \eqref{eq:deterministic}
we get a condition on $\delta t$
\begin{equation*}
    c_k\rho_1+d_k \ll \Abs{(c_k\rho_1-d_k)}^2 \delta
    t\ll\Abs{c_k\rho_1-d_k}k.
\end{equation*}
We can find $\delta t$ which satisfies this condition if 
\begin{equation}
    c_k\rho_1+d_k \ll\Abs{c_k\rho_1-d_k}k.\label{eq:deterministic:con}
\end{equation}
This does not hold for every value of $k$. 
In particular, we notice
that for $k=k_c$ the RHS is zero. 
Hence, the $k(t)$ that can be approximated as deterministic must be
bounded away from $k_c$. 

For $k\gg1$ in \eqref{eq:deterministic:con} we insert the approximations (\ref{eq:d_k}, \ref{eq:detailed:balance}) for $d_k$ and $c_k$, and then examine its asymptotic form as $\eta\goto0^+$. 
The result is 
\begin{equation}
  \brk{2+\eta\brk{1-\brk{\frac{k}{k_c}}^{-\third}}}\ll \eta\Abs{1-\brk{\frac{k}{k_c}}^{-\third}}k.\label{eq:det:con:2}
\end{equation}

For clusters with $k\gg k_c$, this condition is satisfied trivially for $\eta\ll1$. 
For $k$ on the order of $k_c$, we need to dig a little deeper.

Since we are looking at the $\eta\goto0$ limit, we look at the leading order terms of \eqref{eq:det:con:2}
 \begin{equation*}
  \frac2{\eta k}\ll \Abs{1-\brk{\frac{k_c}{k}}^{\third}}.
\end{equation*}
For $k$ near $k_c$ we expand the RHS around $k=k_c$.
Keeping the leading order terms of the Taylor series of the RHS, we
find an explicit condition for the evolution of clusters to be treated
deterministically: $k$ must be bounded away from $k_c$ according to
\begin{equation*}
  \frac6\eta\ll\Abs{k-k_c}.
\end{equation*}
As $\eta\goto0$, the excluded domain grows.
On the other hand, its  size relative to $k_c\sim\brk{\frac{\sig}{\eta}}^3$ shrinks to zero: 
\begin{equation*}
  \frac{6\eta^2}{\sig^3}\ll\Abs{\frac{k-k_c}{k_c}}.
\end{equation*}
Therefore, we see that for small super-saturation the evolution of
clusters can be treated deterministically, for all but a vanishingly small domain around $k_c$. 
\section{Effective Super-Saturation, $\eta$, After Ignition}
\label{sub:sec:eta:eff}
It has previously been shown that an initial condition of pure monomer
leads to a transient `ignition' phase in which quasi-static
densities of sub-critical clusters are created \cite{NBC05}.
Since these sub-critical clusters are made of monomers, after the
ignition transient the monomer density will be lower than the
original value, prior to the ignition. 
We compute the value $\eta_*$ of super-saturation after the formation of the $k<k_c$ quasi-static densities, but before significant depletion  by super-critical clusters.

The starting point if the sum \eqref{eq:conservation} truncated to $k\le k_c$ because the contribution from super-critical clusters is insignificant:
\begin{equation*}
  \rho=\sum_{\makebox[0mm]{$\scriptstyle 1\le k\le k_c$}} k\rho_k.
\end{equation*}
Since the sub-critical cluster densities are quasi-static, we
approximate them by the equilibrium densities \eqref{eq:rho_k:equil}, and the corresponding approximation of the sum is
\begin{equation*}
  \rho=\sum_{\makebox[0mm]{$\scriptstyle 1\le k\le k_c$}} k\rho_1^ke^{-\frac{\eps_k}{k_BT}}.
\end{equation*}
Recalling the definition of the super-saturation,
\eqref{eq:supersaturation},  $\rho_1 = \rho_s (1+\eta_*)$, and taking
the two-term expansion as $\eta_*\goto0^+$, we find
\begin{equation*}
\rho\sim\rho_c+\eta_*\sum_{\makebox[0mm]{$\scriptstyle 1\le k\le k_c$}}k^2\rho_s^ke^{-\frac{\eps_k}{k_BT}}.
\end{equation*}
The sum on the RHS can be obtained by differentiating the series
representation  \eqref{eq:conservation:equil} of $\rho(\rho_1)$, setting $\rho_1=\rho_s$, multiplying by $\rho_s$, and then truncating to $k\le k_c$.
The derivative series converges at $\rho_1=\rho_s$, so for $\eta\goto0$ (and hence for $k_c\goto\infty$), the sum is asymptotic to $\rho_s\rho'(\rho_s)$.
Hence,
\begin{equation*}
  \rho\sim\rho_c+\eta_*\rho_s\rho'(\rho_s), 
\end{equation*}
and so,
\begin{equation*}
  \eta_* = \oneover{\rho'(\rho_s)} \frac{\rho-\rho_c}{\rho_s}.
\end{equation*}

%
%
%
\section{Numerical Solution of Integral Equation}
\label{sec:numerical:explain}
The nucleation era solution required an integral equation to be
solved. 
The scaled equation 
\begin{equation}
  \delta\eta(t)=-\int_0^t \brk{\frac23\brk{t-\phi}}^\frac32
  e^{\delta\eta(t)}\, d\phi
\end{equation}
determines the change in the supersaturation with time. 
We found the approximate solution of this equation on a discrete set
of point separated by a fixed interval $h$. 
The value at each point was calculated from the trapezoidal rule
approximation to the integral of the solution up to time $t+h$. 
This is \emph{not} implicit, since the $t-\phi$ term in the
integrand vanishes at $\phi=t$. The order of accuracy of the method is found  to be $\frac32$
 by numerical experiment. This fractional order is most likely
due to the cusp in the integrand which adds an error of $h^{3/2}$ to
the integral. 

The value R=1.343 was found by using 1000 points in the interval $(0,4)$.
The error in R is estimated to be 0.001, while the error in the
flux itself is estimated to be $10^{-6}$. The errors were obtained
by comparing to the results of the same numerical calculation with 4
times as many points.

\section{Asymptotic Matching Between  Nucleation and Growth Eras}
\label{sec:nuc:grow:match}
Here, we  examine $q(x,t)$  in the overlap
domain $\tau^2\ll t\ll1$  between nucleation and growth given by \eqref{eq:r:as:q:growth}
\begin{equation}
  q(x,t) =\frac{a}{\tau R}\,\brk{\frac{K}{\tau^3}+\frac{ax}{\tau}}^{-\third} j\fbrk{\frac{t}{\tau^2}-\frac32\brk{\frac{K}{\tau^3}+\frac{ax}{\tau}}^\frac23}.\label{eq:r:as:q:growth:app}
\end{equation}
For small $t$, $K(t)$ has asymptotic behavior given by
\eqref{eq:K:small:time}, and $a(t)$ is a simple function of
$K$, as given by \eqref{eq:a:1}
\begin{equation}
  K(t) \sim \brk{\frac23t}^{\frac32},\qquad a = CK^{\frac13}. \label{eq:K_and_a:small:time}
\end{equation}
Selectively substituting the expressions in \eqref{eq:K_and_a:small:time} for
$K(t)$ and $a(t)$ into \eqref{eq:r:as:q:growth:app} yields,
\begin{align*}
  q(x,t) &= \frac{ CK^{\third}}{\tau R}
  \frac{K^{-\third}}{\tau^{-1}}\brk{1+\frac{\tau^2ax}{K}}^{-\third}\,
  j\brk{\frac{t}{\tau^2}-\frac{K^\frac23}{\tau^2}\brk{1+\frac32\frac{\tau^2ax}{K}}^\frac23}.
\intertext{Cancellations and additional use of
    \eqref{eq:K_and_a:small:time} give}
  &= \frac{C}{R}\brk{1+\frac{\tau^2ax}{K}}^{-\third}\,
  j\brk{\frac{t}{\tau^2}\brk{1 - \brk{1+\frac32\frac{\tau^2Cx}{t}}^\frac23}}.
\intertext{Therefore, as $\tau\goto0$, $t$ fixed}
 q(x,t) &=Q(x)\sim \frac{C}{R}\, j\brk{-Cx}.
\end{align*}
\section{Finding $\eta$, the Details}
\label{sub:sec:finding:eta}
Equations (\ref{eq:advection:coarse}--\ref{eq:def:M}) describe  the
evolution of the particle size distribution \emph{implicitly}.
The the supersaturation, $\eta$, is needed for the advection
velocity \eqref{eq:eta:delicate}, but it has no explicitly defining equation.

To find $\eta$, we  derive another equation for $\dot K$ from
\eqref{eq:eta:delicate} and the PDE (\ref{eq:advection:coarse},
\ref{eq:advection:coarsening:scaled})
\begin{equation}
  K = \oneover{M} \quad \Longrightarrow \quad \dot K = -\frac{\dot
  M}{M^2}.\label{eq:K:dot:2:app}
\end{equation}
The value of $\dot M$ can be found from the definition of $M$ 
\eqref{eq:def:M}, the advection equation \eqref{eq:advection:coarse},
and integration by parts
\begin{equation}
  \dot M= \int_0^1 y\, q_t \,dy=\int_0^1 wq\, dy \quad\Longrightarrow\quad \dot
  K = -\oneover{M^2}\int_0^1 wq\, dy .\label{eq:M:dot:app}
\end{equation}
Comparing the two expressions for $\dot K$ in \eqref{eq:K:coarse} and
\eqref{eq:M:dot:app} gives and expression for $\eta$,
\begin{equation}
 \eta = M^\third\brk{1-\frac{\int_0^1 wq\,dy}{M^2}},\label{eq:eta:implicit:app}
\end{equation}
which is  implicit because $w$
\eqref{eq:advection:coarsening:scaled} depends on $\eta$. 
To untangle the dependence on $\eta$, we write $\int_0^1wq\,dy$ using $\eta$ and various moments of $q$.  
\begin{align}
\int_0^1wq\,dy&= K^{-\frac23}\eta\int_0^1\brk{y^\third-y} q\,dy  + K^{-1}\int_0^1 (y-1)q\, dy\notag\\
&= K^{-\frac23}\eta\brk{M_{\third}-M}+K^{-1}(M-M_0).\label{eq:int:advection:app}
\end{align}
Where $M$, $M_0$ and $M_\third$ are moments of the distribution $q$,
\begin{equation*}
M =   \int_0^1 y  q\,dy,\qquad M_0 = \int_0^1 q\,dy,\qquad  M_\third = \int_0^1 y^\third q\,dy.
\end{equation*}

Substituting \eqref{eq:eta:delicate} and \eqref{eq:int:advection:app} in
\eqref{eq:eta:implicit:app} for $\eta$ results in 
\begin{equation*} 
  \eta = M^\third\brk{1-\frac{M^\frac23 \eta\brk{M_\third-M}+M(M-M_0)}{M^2}}\notag,
\end{equation*}
which can be massaged into a much simpler expression for $\eta$:
\begin{equation*}
  \eta = \frac{M^\third M_0}{M_\third}\label{eq:eta:simple:app}.
\end{equation*}
Hence, we end up with the following advection PDE to solve:
\begin{align*}
  0&=q_t+(wq)_y,\\
  w &= M\brk{\frac{M_0}{M_{\third}}\brk{y^\third-y}+(y-1)}.
\end{align*}

\end{document}

%% file: MyPackages.tex
\usepackage{version}
\usepackage{amsmath}
\usepackage{graphicx}
\usepackage{amsthm}
\usepackage{amssymb}
\usepackage{alltt}
\usepackage{epic}
\usepackage{eepic}
\usepackage{shadow}
\usepackage{array}
\usepackage{enumerate}
\usepackage[normalem]{ulem}
\usepackage{epsfig}
\usepackage{multicol}
\usepackage{ifthen}
\usepackage{bbold}

%% file: MyMacros.tex
\newcommand{\brk}[1]{\left(#1\right)}          
\newcommand{\fbrk}[1]{\!\left(#1\right)}       
\newcommand{\Brk}[1]{\left[#1\right]}          
\newcommand{\BRK}[1]{\left\{#1\right\}}        

\newcommand{\Average}[1]{\left<#1\right>}      

\newcommand{\Abs}[1]{\left| #1 \right|}        

\newcommand{\goto}{\rightarrow}

\newcommand{\sups}[1]{{${}^{\text{#1}}$}} 
\newcommand{\subs}[1]{{${}_{\text{#1}}$}} 

 {

\newtheorem*{prop*}{Proposition}
 }

\newcounter{sect}

\newcommand{\oneover}[1]{{\frac{1}{#1}}}

\newcommand{\third}{\frac{1}{3}}

\newcommand{\R}{\mathbb R}

\renewcommand{\O}{{\mathcal O}}

\newcommand{\sig}{\sigma}
\renewcommand{\phi}{\varphi}

\newcommand{\eps}{\varepsilon}

\def\clap#1{\hbox to 0pt{\hss#1\hss}}